\documentclass[12pt,preprint]{aastex}
\def\Ho {H_{0}}

\begin{document}

\title{THE MEGAMASER COSMOLOGY PROJECT. 
\uppercase\expandafter{\romannumeral9}.\\
BLACK HOLE MASSES FOR THREE MASER GALAXIES
}
\author{F. Gao\altaffilmark{1,2,3}, J. A. Braatz\altaffilmark{2},  
        M. J. Reid\altaffilmark{4},   
        J. J. Condon\altaffilmark{2}, J. E. Greene\altaffilmark{5}, 
            C. Henkel\altaffilmark{6,7}, C. M. V. Impellizzeri\altaffilmark{2,9},
            K. Y. Lo\altaffilmark{2},  C. Y. Kuo\altaffilmark{8}, 
 D. W. Pesce\altaffilmark{10}, 
J. Wagner\altaffilmark{6,11} and W. Zhao\altaffilmark{1}
} 
            
\altaffiltext{1}{Key Laboratory for Research in Galaxies and Cosmology,
 Shanghai Astronomical Observatory, Chinese Academy of Science,
  Shanghai 200030, China}
\altaffiltext{2}{National Radio Astronomy Observatory, 
520 Edgemont Road, Charlottesville, VA 22903, USA}
\altaffiltext{3}{Graduate School of the Chinese Academy of Sciences,
 Beijing 100039, China}
\altaffiltext{4}{Harvard-Smithsonian Center for Astrophysics, 60 Garden Street,
 Cambridge, MA 02138, USA}
\altaffiltext{5}{Department of Astrophysics, Princeton University,
 Princeton, NJ 08544, USA }
\altaffiltext{6}{Max-Planck-Institut f\"ur Radioastronomie,
 Auf dem H\"ugel 69, 53121 Bonn, Germany}
\altaffiltext{7}{King Abdulaziz University, P.O. Box 80203,
 Jeddah 21589, Saudi Arabia}
\altaffiltext{8}{Department of Physics, National Sun Yat-Sen University,
 No.70, Lianhai Rd., Gushan Dist., Kaohsiung City 804, Taiwan (R.O.C.)}
\altaffiltext{9}{Joint Alma Office, Alsonso de Cordova 3107,
 Vitacura, Santiago, Chile }
\altaffiltext{10}{Department of Astronomy, University of Virginia,
 P.O. Box 400325, Charlottesville, VA 22904, USA }
\altaffiltext{11}{Korea Astronomy and Space Science Institute,
 776, Daedeokdae-ro, Yuseong-gu, 305-348, Daejeon, Republic of Korea}

\begin{abstract}
As part of the Megamaser Cosmology Project (MCP), we present VLBI maps 
of nuclear water masers toward five galaxies.  The masers originate in
sub-parsec circumnuclear disks.  For three of the galaxies, we
fit Keplerian rotation curves to estimate their supermassive black hole (SMBH) masses, and
determine (2.9 $\pm$ 0.3) $\times~10^{6}M_\odot$ for J0437+2456, (1.7 $\pm$ 0.1)
$\times~10^{7}M_\odot$ for ESO 558$-$G009, and (1.1 $\pm$ 0.2) $\times~10^{7}M_\odot$
for NGC 5495.  In the other two galaxies, Mrk 1029 and NGC 1320, 
the geometry and dynamics are more complicated and preclude robust black hole mass estimates.
Including our new results, we compiled a list of 15 VLBI-confirmed
disk maser galaxies with robust SMBH mass measurements. With this sample, 
we confirm the empirical relation of $R_{out} \propto 0.3 M_{SMBH}$ reported in Wardle \& Yusef-Zadeh (2012). 
We also find a tentative correlation between maser disk outer radii and WISE
luminosity. 
We find no correlations of maser disk size with X-ray 2-10 keV luminosity or [O III] luminosity.  

\end{abstract}

\section{INTRODUCTION}

    The mass of a supermassive black hole (SMBH) is a fundamental parameter of galaxies. Among 
the dynamical tracers used to measure the SMBH mass directly (e.g. from gas or stars), 22 GHz 
H$_{2}$O megamasers provide the most precise mass estimates outside of our Galaxy. As first demonstrated 
in the archetypal maser galaxy NGC 4258 (Greenhill et al. 1995; Miyoshi et al. 1995), maser features in such systems trace
 an edge-on, thin Keplerian disk on sub-parsec scales around the central SMBH and the enclosed mass
  can be constrained to $\pm3$\% (Humphreys et al. 2013). Such disk masers are usually detected in Seyfert 2
   galaxies, where the nuclear disk is edge-on and maser emission is beamed into our line-of-sight (LOS).     
    
    Over the last two decades, disk-maser galaxies have emerged as an important sample for 
    high-precision measurement of SMBH mass, thanks to large surveys 
    (e.g. Braatz et al. 1996; 1997; Greenhill et al. 2003; Braatz et al. 2004; Kondratko et al. 2006; 
    Braatz \& Gugliucci 2008). In particular, new maser disks are being
discovered and mapped by the Megamaser Cosmology Project (MCP, Reid et al. 2009; Braatz et al. 2010)
    \footnote{Project webpage: https://safe.nrao.edu/wiki/bin/view/Main/MegamaserCosmologyProject}, 
    which ultimately aims to use the maser disks to determine the Hubble constant $\Ho$ to better than 3\% uncertainty.
    So far, the MCP has published 7 high-accuracy SMBH masses (Kuo et al. 2011). Including
    results from other groups (Herrnstein et al. 1999; Greenhill et al. 1997, 2003; Kondratko et al. 2005, 2008;
    Mamyoda et al. 2009; Ishihara et al. 2001), there are 16 sources measured altogether. 
    
   The SMBH mass results can be used to investigate the M $-$ $\sigma$ relation, as has been done by 
   Greene et al. (2010; 2016). The masses derived from maser galaxies are between 10$^{6}$ 
   and 10$^{7.5} M_\odot$ and occupy the less-explored lower-mass end of the M $-$ $\sigma$ plot. 
   Currently maser galaxies fall below the empirical M $-$ $\sigma$ relation (e.g. McConnell \& Ma 2013). 
   As discussed in Greene et al. (2010) and listed in Kormendy \& Ho (2013), many of the host galaxies
   of maser disks have pseudo-bulges rather than classical bulges, suggesting that galaxies with
    pseudo-bulges may follow a different
   M $-$ $\sigma$ relation than galaxies with classical bulges. As the disk maser sample grows, we could 
   better test whether disk masers are always seen in systems with 10$^{6}$ to 10$^{7.5} M_\odot$ SMBHs and
   whether maser galaxies with classical bulges follow the empirical M $-$ $\sigma$ relation. 
    
    Some galaxies show maser profiles that differ from the ``classical" triple sets of lines characteristic
    of maser disks. These more complicated systems may harbor outflows
    (e.g. Circinus; Greenhill et al. 2003; NGC 3079; Kondratko et al. 2005;) and 
    jets (e.g. NGC 1068; Greenhill et al. 1997). Though they are less studied compared to
    disk masers, as more are imaged, they may become an important sub-sample to
    study the outflows and interactions of the central AGN with its environment.        
    
    Here we present high-resolution VLBI images and rotation curves of
    5 new maser galaxies: J0437+2456, ESO 558$-$G009, NGC 5495,
    Mrk 1029, and NGC 1320.  The first three have spectral profiles indicative of ``classical'' Keplerian disk masers,
    while the last two have profiles suggesting more complicated nuclear structure (e.g. Pesce et al. 2015).
    We note that J0437+2456 and ESO 558$-$G009 are also
    current targets for distance measurements with more extensive MCP observations.  Here we
    present preliminary VLBI maps and SMBH masses,
    based on fitting the rotation curve of high-velocity maser features.
    We describe our observations
    and data reduction in Section 2. Then Section 3 shows the VLBI images and the rotation curves,
    followed by our analysis of the SMBH masses on each source. Using these results, in Section 4, 
    we define a sample of 15 ``clean'' disk masers and explore the relation between maser disk radius, 
    SMBH mass and AGN luminosities. Then we give our conclusions in Section 5.

\section{OBSERVATIONS}
	
	\subsection{New maser discoveries}
	
We present the first maser maps for 5 galaxies here. Three of the
masers (J0437+2456, ESO 558$-$G009, Mrk 1029) were discovered in 2010
by our MCP survey. This survey covers Sy 2 galaxies identified from
the Sloan Digital Sky Survey (SDSS), Sy 1.8 $-$ 2.0 from NASA/IPAC
Extragalactic Database (NED), and Sy 2 galaxies identified from the
Two Micron Redshift Survey (2MRS). All survey observations were
conducted using the Robert C. Byrd Green Bank Telescope
(GBT)\footnote{The Green Bank Telescope is a facility of the National
  Radio Astronomy Observatory.}.  The survey will be described fully in a
forthcoming paper (Braatz et al. in prep).

The maser in NGC 5495 was discovered by Kondratko et al. (2006) during
a survey of Sy 1.8 $-$ 2.0 and LINER galaxies with $v_{sys} \le$ 14000
km s$^{-1}$ using the NASA DSN antennas, and the maser in NGC 1320 was
discovered by Greenhill et al. (2009) in a survey of galaxies with
inclined stellar disks with the GBT. 
    
    All of these galaxies are Sy 2 AGN, with recession
    velocities between 2663 km s$^{-1}$ and 9076 km s$^{-1}$. Their
    host galaxies are all late-type galaxies. We list their host
    galaxy properties in
    Table~\ref{table:positions}. Fig.~\ref{fig:GBTspec} shows a GBT
    single-dish spectrum for each source observed closest to the VLBI
    observing dates (with a maximum time separation of 60 days). 
    For NGC 1320, we show the GBT spectra observed before 
    and after the VLBI observation in
    Fig.~\ref{fig:GBTspec}, showing significant variability.      
    The original spectral resolution of the GBT
    spectra was 24.4 kHz ($\approx0.3$ km s$^{-1}$).  However, for
    comparison with the VLBI results, these spectra were
    smoothed to the same spectral resolution as the
    VLBI observations.

	\subsection{VLBI observations}

    The VLBI observations were conducted between 2010 and 2014 using
    the Very Long Baseline Array(VLBA)\footnote{The VLBA is a facility of
    the National Radio Astronomy Observatory, which is operated by the
    Associated Universities, Inc. under a cooperative agreement with the
    National Science Foundation (NSF).} together with the GBT, and in 
    some cases also with the Effelsberg 100m telescope (EB) and the
    phased Karl G. Jansky Very Large Array (VLA). We list the basic VLBI
    observing information in Table~\ref{table:observation}, including the 
    experiment code, date observed, track duration, antennas used, beam
    size, sensitivity, and the observing strategy.  All but one (J0437+2456) 
    of the galaxies were observed with similar setups as described below.
    
\begin{figure}
\epsscale{0.80}
\plotone{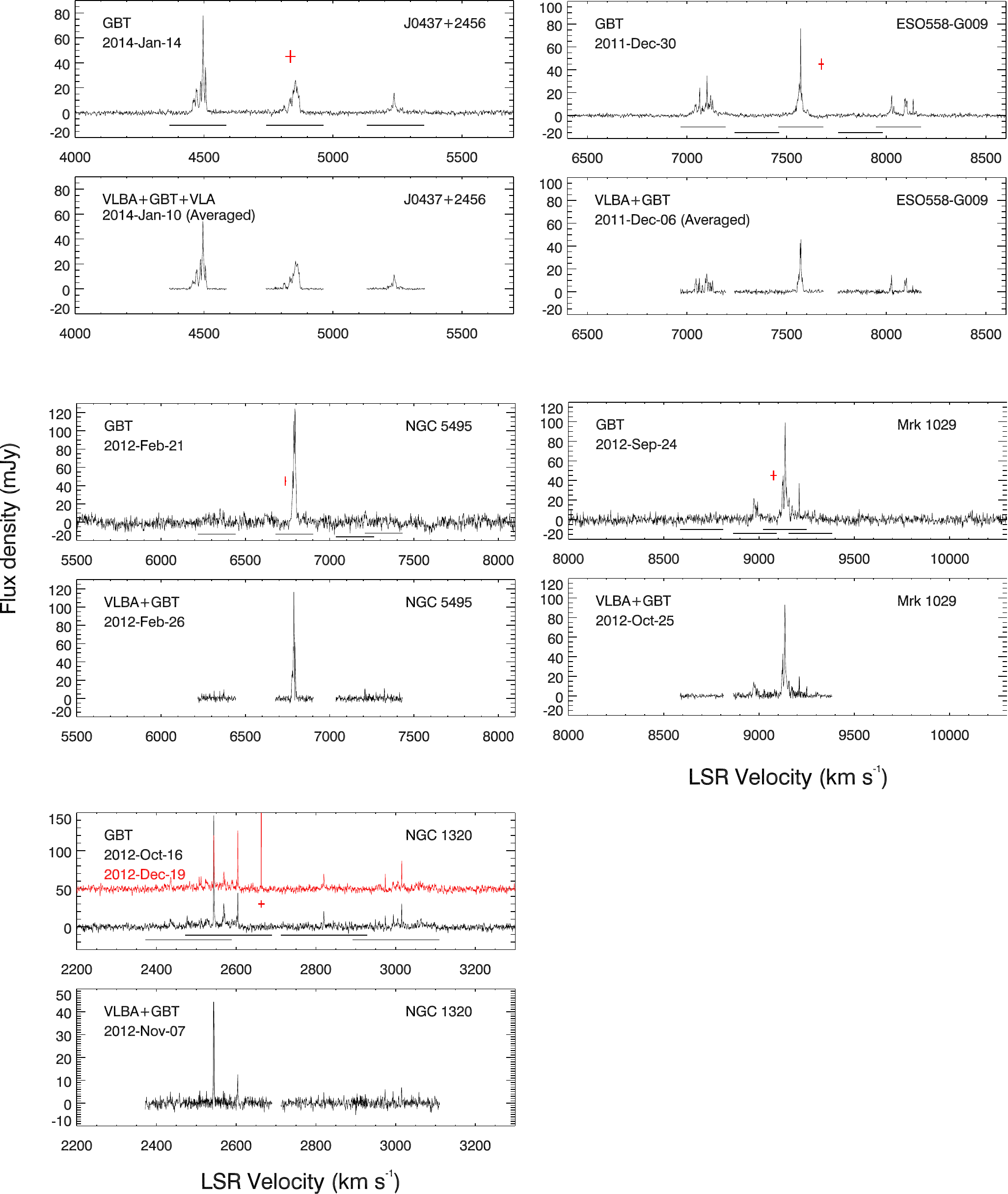} 
\caption{A GBT single-dish spectrum and the VLBI spectrum for each of the 5 targets. For each
galaxy, the single-dish spectrum is shown in the upper panel. Except for NGC 1320, we use the
same scale for intensity when comparing the two spectra. The single-dish spectrum
is smoothed to have the same channel size as the VLBI spectrum, except for ESO 558$-$G009,
in which the VLBI spectral resolution for the systemic part is $\approx$ 3.55 km~s$^{-1}$, while the
spectral resolution in the single-dish spectrum is 1.8 km~s$^{-1}$. The mismatched channels cause the
apparent difference in intensity of the systemic features between the single-dish spectrum and the
VLBI spectrum. The red crosses indicate the recession velocity of the galaxy reported from NED, with the
horizontal bar indicating the uncertainty. The black solid horizontal lines beneath the spectra
indicate the velocity coverage in our VLBI imaging observations. For J0437+2456, the solid
lines show the zoom-band we used for the final correlation pass. The original dual 128 MHz bands
cover the full velocity extent for J0437+2456 as shown here. 
 \label{fig:GBTspec}
        }
\end{figure}

    For ESO 558$-$G009, Mrk 1029, NGC 5495 and NGC 1320, we conducted
    the observations using a ``self-calibration'' strategy, in which we remove 
    atmospheric phase fluctuations using a strong, narrow maser line from the target maser galaxy. 
    Two ``geodetic blocks,''
    bracketing the maser observations were used to calibrate the tropospheric delay
    and clock errors at each antenna. The geodetic blocks were generated by
    SCHED automatically, with each block lasting about 45 minutes. For the
    details of geodetic block setup and data reduction, please refer to
    Mioduszewski et al. (2000) and Reid et al. (2009). For the maser observations,
    we observed 4 IF pairs in dual polarization, with a bandwidth of 16 MHz
    in each IF. Three IF pairs were set to cover the systemic, the blue-shifted
    and the red-shifted masers respectively, with the fourth covering any additional
    emission seen in the single-dish spectrum. The spectral coverage in our
    VLBI observations is shown in Fig.~\ref{fig:GBTspec} as the horizontal
    solid lines below the GBT spectra. We observed a nearby delay calibrator
     about every 50 minutes to calibrate the single band
     delay, and a bandpass/fringe-finder source was observed twice in each
     track. Our data were recorded with the Mark 5B recording system
     (a.k.a the VLBA legacy system) with the maximum recording rate
     of 512 Mbit~s$^{-1}$ and maximum bandwidth of each baseband limited to 16 MHz.

     For ``self-calibration" observations, to map the maser distribution optimally
     and derive the best SMBH mass, an accurate absolute position of the reference feature is
     needed. A large offset between the true absolute position
     and the position used in VLBI correlation will introduce a significant phase error that
     varies with observed frequency, contributing an additional positional uncertainty for each
     maser spot, distorting our final map and affecting the measured warping in the
     position angle (P.A.) direction (Gao et al. 2016). Typically an absolute position accuracy 
     of $\approx$ 10 mas is sufficient in our MCP observations. Otherwise the systematic errors
     resulting from this effect should also be considered (see Section 3.3 for more details).          
    
    We conducted a phase-referenced VLBI
    observation of J0437+2456 to measure the absolute position of the maser.
    We observed an external calibrator J0435+2532 (0.70$^{\circ}$ away)
    every 70 seconds to remove the atmospheric phase variations. We followed
    the data reduction process for phase-referencing observations as described
    in Reid et al. (2009). Finally we derived the sky position of the peak maser
    emission at velocity 4495.38 km~s$^{-1}$, to be at RA= $04^h37^m03^s.6840$
    and DEC= $+24^{\circ}$56'06".837, with uncertainties of 1.3 mas and 2 mas
    respectively. We used this position for data correlation in the subsequent ``self-cal'' tracks. 
    
    For ESO 558$-$G009, we used the VLA in B configuration to measure
    the maser position RA= $07^h04^m21^s.011$ and DEC=
    $-21^{\circ}$35'19".948 with uncertainties of 33 mas and 64 mas,
    respectively, prior to our VLBI observations. To better constrain
    the maser position, we conducted a phase-referenced VLBI
    observation (BB313AF) in which an external calibrator J0702-1951
    (1.77$^{\circ}$ away) was observed every other minute. However,
    due to the relative weakness of the peak maser line intensity ($<$
    40 mJy, measured from the GBT single dish spectrum taken within
    one month to the VLBI observation), and phase de-coherence in the
    phase transfer from J0702-1951 to ESO 558$-$G009, we were unable to
    detect the signal from ESO 558$-$G009. So we applied the VLA
    position of ESO 558$-$G009 for VLBI correlation.

    Because of the relatively low intensity of maser lines in J0437+2456 (with peak
    intensity usually less than 50 mJy), the majority of ``self-cal'' tracks for
    J0437+2456 were conducted between 2013 and 2014 (with 3 tracks presented
    here), when we could include the phased-VLA in our observation array,
    boosting our sensitivity by $\approx40$\%.  During observations, we 
    ``phased-up'' all of the VLA antennas prior to each VLBI scan.  This was done by
    observing J0435+2532 ($>$ 140 mJy, 0.7$^{\circ}$ away from J0437+2456)
    for 1 min. To minimize decorrelation, we set the maximum
    VLBI scan length to be 10 mins with the VLA in B- or C-configuration.
    We also added a flux calibration scan for the VLA to get the combined system
     temperature for the phased array. We used the Mark 5C
     recording system, which offers a maximum recording rate of 2 Gbit~s$^{-1}$,
     enabling us to cover the entire maser spectrum with two
     128 MHz bands ($\approx1700$ km s$^{-1}$ for each band) in dual polarization
      and 2-bit sampling.  In order to obtain high spectral resolution for the maser data, 
      we re-correlated 16 MHz portions of these broad bands with
      the new zoom-band mode offered in the DiFX software correlator (Deller et al. 2007).
      Fig.~\ref{fig:GBTspec} shows the velocity coverage of the zoom-bands for
      J0437+2456 as horizontal dashed lines below the GBT spectrum. For comparison,
       the original dual 128 MHz bands cover the full velocity range shown in Fig.~\ref{fig:GBTspec}.
        For each of the self-calibrated tracks, we did not observe geodetic blocks, since we used
        the hourly delay calibrator observations to solve for and remove residual multi-band delays.
        
        All data were calibrated using the NRAO Astronomical Image
        Processing System (AIPS). For the self-calibration tracks, we
        followed the data reduction process described in
        Reid et al. (2009). The typical solution interval used
        in CALIB varied between 1 and 2 minutes,
        depending on the coherence times in each track.

    For J0437+2456, before the final imaging step, 
    we averaged the (\emph{u,v}) data to 1.4 km s$^{-1}$ per channel
    (100 kHz) to match the typical spectral linewidths from the single-dish spectrum. After
    we made images for tracks BB321X0, BB321X1 and BB321X4 separately, we combined the data
    before imaging, assuming that the strongest feature at
    4496.15 km s$^{-1}$ didn't change its position. Since we observed the tracks within
    a month, this is a safe assumption.      
        
    For ESO 558$-$G009, we averaged the interferometer spectra to 3.55 km
    s$^{-1}$ per channel (250 kHz) for the systemic features, and 1.8
    km s$^{-1}$ per channel (125 kHz) for the high-velocity
    features. Then we calibrated the data from tracks BB278X and
    BB278T separately and combined them in the (\emph{u,v}) domain by
    aligning the strongest systemic feature at velocity 7572 km
    s$^{-1}$. Then we made images from the combined data.
    
    For Mrk 1029 and NGC 5495, we averaged the (\emph{u,v}) data to 1.8 km s$^{-1}$
     per channel (125 kHz). For NGC 1320, no averaging was performed on the
      (\emph{u,v}) data. The raw data has spectral resolution of 0.86 km s$^{-1}$ per channel (62.5 kHz).
      We list the beam size and 1$\sigma$ image sensitivity for each track in Table~\ref{table:observation}.
      The VLBI-imaged maser components for each galaxy are listed in Table~\ref{table:rawdata}.

\section{BLACK HOLE MASS FITTING AND RESULTS}

	\subsection{Fitting Method}
	
    The SMBH mass from a disk maser galaxy can be determined using the rotation curve measured from 
    high-velocity maser features in a VLBI map (e.g. Kuo et al. 2011). The model can be generalized to 3D if
    spectral monitoring observations are made to measure the line-of-sight accelerations of individual maser features,
    which are essential to derive the physical radius of the systemic maser features and thus determine the geometric distance
    to the maser disk (e.g. Humphreys et al. 2013; Reid et al. 2013; Gao et al. 2016).
    Among the 5 galaxies present here, J0437+2456 and ESO558$-$G009 have been observed with the GBT to monitor 
    the spectral features. Analysis of these observations is ongoing, so in this paper we utilize only the rotation curve from 
    high-velocity features to estimate the SMBH masses for all 5 galaxies. A full 3D analysis of the maser disk in 
    J0437+2456 and ESO 558$-$G009 will be present in subsequent work.  

    Our estimation of the SMBH mass depends directly on the distance to the host galaxy.  In
    this paper, we derive the galaxy distances based on their recession velocity with an $\Ho$
    of 73 km~s$^{-1}$ Mpc$^{-1}$, in concordance with the previous MCP work by Kuo et al. (2011).   
    We present all the VLBI images together with the position-velocity (P-V) diagrams and residuals
    in Fig.~\ref{fig:VLBImap} and Fig.~\ref{fig:VLBImap2}. We determine the black hole mass by 
    applying the following steps.

    First we defined the position angle of the disk with respect to the red-shifted masers, from due north to the east.
    According to such position angle, we rotate the VLBI image with respect to a reference feature so that the
    major axis of the maser disk lies horizontally. Then we minimize the vertical offsets of the high-velocity features by
    shifting all the data vertically. We estimate the disk inclination angle as 
    $cos^{-1}i = \bar{\theta_{y}}/\bar{\theta_{r}}$, following the same treatment in Kuo et al. (2011), 
    in which $\bar{\theta_{y}}$ is the averaged 
    vertical offset of the systemic masers from the disk plane defined by the high-velocity masers, $\bar{\theta_{r}}$
    is the averaged angular separation between the systemic masers and the high-velocity masers, which we equate to the 
    radius of the high-velocity features. Here we have assumed all the systemic masers are located on the same 
    radius as the high-velocity masers.
    Then we define the impact parameter in the P-V diagram
    as the radial offset $r_{i} = [(x_{i}-x_{0})^{2}+(y_{i}-y_{0})^{2}]^{\frac{1}{2}}$
    of each maser spot from the reference feature ($x_{0}, y_{0}$).
    Here, $x_{i}$ and $y_{i}$ represent maser positions after the coordinate rotation
    and vertical displacement.  Features on
    the red-shifted side have positive values of the impact parameter, while features on the blue-shifted
    side have negative values. The reference feature is usually the one with a velocity
    closest to the reported recession velocity. If all the systemic
    features are offset from the reported recession velocity, then we use the strongest systemic
    feature as the reference, as in the case of ESO 558$-$G009.

    We fit the rotation curve of the high-velocity features with a Keplerian rotation model, in which 
    
    \begin{equation}
    {v_{\rm{rot_{i}}}}=[{\frac{GM}{(r_{i}-r_{0})}}]^{\frac{1}{2}}. \label{eqn:eq1}
    \end{equation}
    
    \noindent Here $v_{\rm{rot_{i}}}$ is the Keplerian rotation speed of the $i$th maser feature around a
    central black hole with mass $M$. The reference position $r_{0}$ refers to the dynamical center on the maser
    disk plane, and $r_{i}$ is the impact parameter of the $i$th maser feature with respect to this reference
    position. The observed recession velocity is related to the Keplerian rotation velocity by the following:
    
    \begin{equation}
    v_{\rm{LSR,opt}}=v_{\rm{rot_{i}}} {\rm sin}~i+v_{0}+v_{\rm{rel_{i}}} \label{eqn:eq2}
    \end{equation}
    
    \noindent where
    
    \begin{equation}
    v_{\rm{rel_{i}}}\simeq\frac{1}{2c}[(v_{\rm{rot_{i}}}+v_{0})^{2}+v_{\rm{rot_{i}}}^{2}{\rm cos^{2}}~i+2v_{\rm{rot_{i}}}^{2}]. \label{eqn:eq3}
    \end{equation}
    
    \noindent Here $v_{\rm{rel_{i}}}$ is the velocity correction for both the special and general relativistic effects
    for the $i$th maser feature (Herrnstein et al. 2005), and $v_{0}$ is the systemic velocity of the
    galaxy, defined in the relativistic frame. The $v_{\rm{LSR,opt}}$ is the LSR velocity
    of each observed maser feature, using the optical definition of velocity. $i$ is the inclination
    angle of the maser disk derived above, with 90$^{\circ}$ corresponding to edge on. Here we assume all the
    high-velocity features are located on the mid-line of the disk, where the mid-line is defined as the 
    diameter through the disk perpendicular to the LOS. 
    
     A Bayesian method was used in our fitting to get the best-fit $M$, $r_{0}$, and $v_{0}$. We use the Markov
     Chain Monte-Carlo (MCMC) approach to sample the parameter space and the Metropolis-Hastings algorithm
     to explore the probability density function (PDF) for each parameter, in which we aimed at 25\% acceptance rate.
     The $\chi^{2}$ of the model fit is given by
    
    \begin{equation}
    \chi^{2}=\sum\limits_{i} \frac{(v_{\rm{LSR,opt,i}}-v_{\rm{rot_{i}}}-v_{0}-v_{\rm{rel_{i}}})^{2}}{\sigma_{v,i}^{2}}. \label{eqn:eq4}
    \end{equation}
    
    \noindent Here the velocity uncertainties $\sigma_{v,i}$ are expressed by the uncertainties
    in the measured position of masers:  
    
    \begin{equation}
    \sigma_{v,i}^{2}=\frac{1}{4}v_{\rm{rot_{i}}}^{2}\frac{(x_{i}-x_{0})^{2}\sigma_{x,i}^{2}+(y_{i}-y_{0})^{2}\sigma_{y,i}^{2}}{[(x_{i}-x_{0})^{2}+(y_{i}-y_{0})^{2}]^{2}} \label{eqn:eq5}
    \end{equation}
    
    \noindent where $\sigma_{x,i}$ and $\sigma_{y,i}$ are the measured VLBI positional uncertainties projected
    along and perpendicular to the maser disk plane after the coordinate rotation.
    Note that values of $y_{0}$ are fixed
    to zero in the actual fitting since we place the dynamical center on the maser disk plane, and assume
    the maser disk is flat with all high-velocity features on the mid-line of the disk. 
    As shown in Equation \ref{eqn:eq2}, 
    there is a degeneracy between the inclination angle and the rotation velocity. Adding the inclination angle
    as another free parameter in the fitting will dramatically slow down the fitting convergence, so we didn't 
    fit the inclination angle here. 
        
    We use flat prior for the 3 fitting parameters, unless otherwise stated. Typically, we discarded 10$^{7}$ 
    burn-in trials before generating 9$\times$10$^{7}$ trials for parameter estimation. 
    We binned these 9$\times$10$^{7}$ trials to generate the final PDF. For each parameter,
     we quote the mode in the final PDF as the fitted value, and sum up the final PDF to find the $\pm$ 34\% 
     from the mode as the 1$\sigma$ uncertainty. All uncertainties have been scaled by $\sqrt{\chi^{2}/N}$.  
    Below we describe the VLBI and black hole mass fitting results for each galaxy individually, and in Section 3.2 
    we list several sources of systematic errors in the fitting process. We note that the SMBH mass uncertainties
    reported in Section 3.2 do not include the uncertainty in $H_{0}$, but they are added later as part of the systematics.

	\subsection{Notes on individual sources}
	
		\subsubsection{J0437+2456}
    The VLBI map of J0437+2456 shown in Fig.~\ref{fig:VLBImap} has a flux cutoff of 1.7 mJy
    beam$^{-1}$ channel$^{-1}$ (with a channel width of 100 kHz), which is at the 5$\sigma$ level. 
    We highlight features above 5 mJy in filled circles.
    The VLBI map shows an edge-on disk in a linear configuration, with a P.A. of 20$^{\circ}$ $\pm$ 2$^{\circ}$. 
    There is a slight misalignment between the systemic masers and the high-velocity masers. 
    The estimated inclination angle of the disk is 81$^{\circ}$ $\pm$ 1$^{\circ}$.

    The best-fit (with reduced $\chi^{2}$ of 1.07) black hole 
    mass is (2.9 $\pm$ 0.1) $\times~10^{6}M_\odot$, with recession velocity V$^{\rm{LSR, opt}}_{0}$ of 4809.5 $\pm$ 10.5 km s$^{-1}$.
    The dynamical center is located + 0.12 mas from the original reference position (with velocity of 4835 km s$^{-1}$)
     on the major axis of the maser disk.
    The inner and outer radius of the disk are about 0.11 and 0.42 mas, which corresponds to 0.035
    and 0.13 pc. The uncertainty on the SMBH mass is $\pm$ 3.4\%.    
     We note that on the P-V diagram, the red-shifted features are scattered horizontally with 
     a positive slope on the residual plot, in contrast to the well-fit blue-shifted features. 
    The pattern in the residuals may be influenced by unmodeled disk structure, with the model fit limited
    by low signal to noise. Higher-sensitivity VLBI observations and 3D 
    modeling of the maser disk should resolve the full disk structure. 
    Nevertheless, since it is the blue-shifted masers that mostly constrain the SMBH mass result here 
    (with higher SNR, larger velocity coverage), our best-fit SMBH mass is still robust.
    We show the probability density function for the dynamical center position, 
    the SMBH mass and the recession velocity in Fig.~\ref{fig:mcmcpdf}.
        
    From single-dish monitoring observations aimed at measuring the accelerations of the
    systemic features, and unrelated to the SMBH mass measurement we present here, we found the
    strongest features in J0437+2456 increased in flux from 45 mJy
    (2013 December 19) to 85 mJy (2014 January 14) then fell back to
    40 mJy (2014 February 19).  By catching such maser brightening in our VLBI
    observations (BB321X0, BB321X1 and BB321X4), we were able to calibrate the interferometer 
    phase on the strongest maser line. 
    Even when including the VLA,
    it is difficult to solve for the fringe phase from a spectral line feature less
    than 50 mJy, with a channel size of 1.4 km s$^{-1}$ and a solution interval of 1-2 minutes.

		\subsubsection{ESO 558$-$G009}
\begin{figure}
\epsscale{0.75}
\plotone{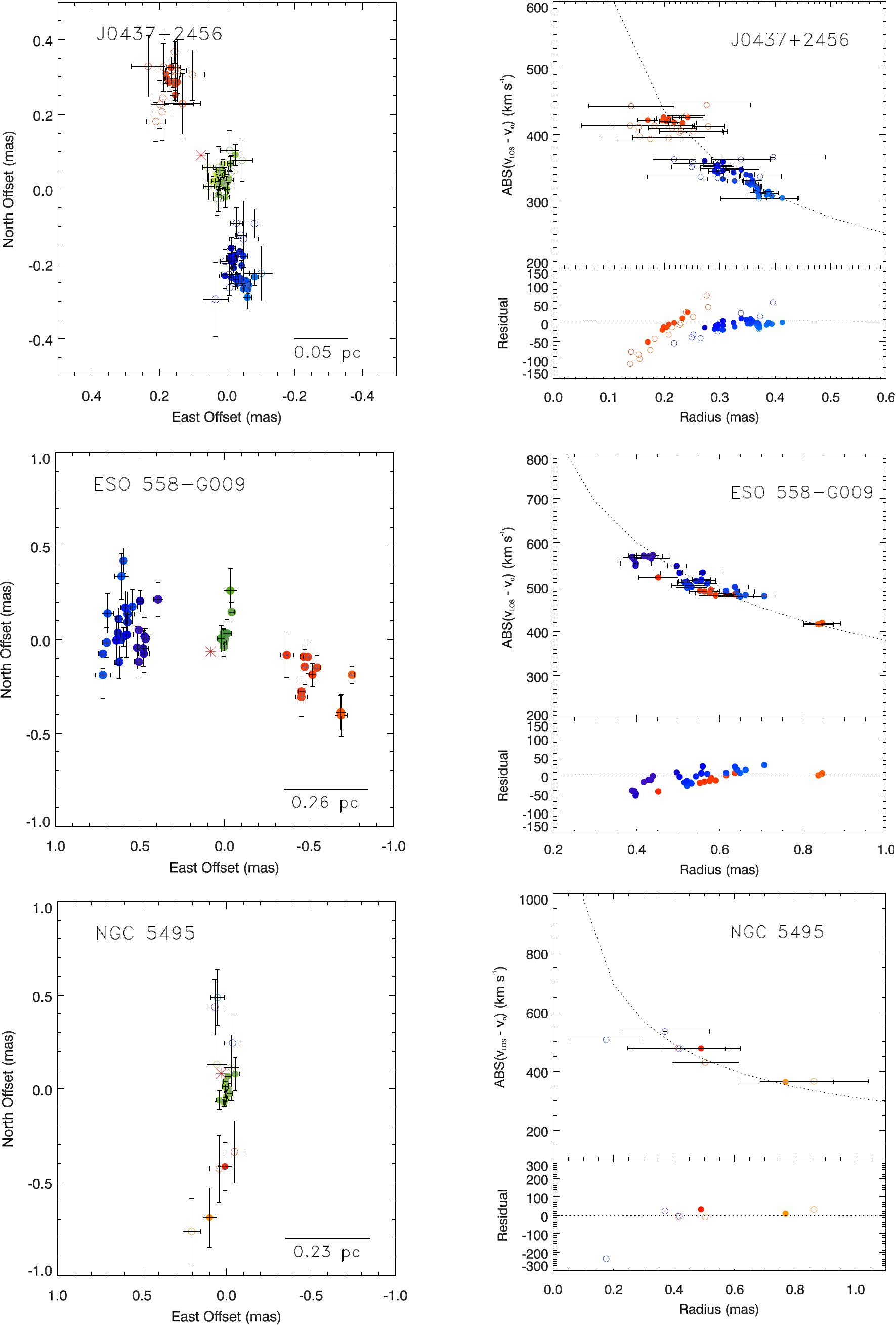} 
\caption{The VLBI maps with P-V diagrams for the 3 disk masers: J0437+2456, ESO 558$-$G009
and NGC 5495. Residuals are shown on the bottom of the P-V diagram. For J0437+2456 and ESO 558$-$G009, 
maser spots with intensity above 5$\sigma$ are shown here, and we highlight maser features brighter than 5 mJy 
in filled circle in J0437+2456. For NGC 5495, features between 3$\sigma$ and 5$\sigma$ are shown in open circle. 
The red asterisk marks the fitted dynamical center on the VLBI map. The dotted
line on the P-V diagram shows the best-fit Keplerian rotation curve. Maser spots on the P-V diagram are referenced
to the fitted dynamical center on the X-axis.
 \label{fig:VLBImap}
        }
\end{figure}

    The 1$\sigma$ noise of the ESO 558$-$G009 VLBI map is about 1.1 mJy
    beam$^{-1}$ channel$^{-1}$, using a channel width of 125 kHz. We plot all
    the maser features brighter than 5.5 mJy in Fig.~\ref{fig:VLBImap}. The map shows
    an edge-on maser disk at a P.A. of 256$^{\circ}$ $\pm$ 2$^{\circ}$. Because of the low
    declination of ESO 558$-$G009 ($-$21$^{\circ}$), the synthesized beam is elongated
    in the north-south direction, approximately perpendicular to the disk, so the map
    does not tightly constrain the disk thickness. 
    The vertical offset of the major axis of the maser disk is -0.082 mas.
    We estimate the inclination angle of the disk to be  98$^{\circ}$ $\pm$ 1$^{\circ}$.
    
     The high-velocity masers in ESO 558$-$G009 are well fit with a Keplerian 
     rotation curve, as shown in Fig.~\ref{fig:VLBImap}.
     The best-fit (with reduced $\chi^{2}$ of 1.28) black hole mass is
     (1.70 $\pm$ 0.02) $\times~10^{7}M_\odot$ and the recession velocity V$^{\rm{LSR, opt}}_{0}$
     is 7595.9 $\pm$ 14.0 km s$^{-1}$. The dynamical center is located -0.064 mas
     from the original reference position (with velocity of 7591 km s$^{-1}$)
     on the major axis of the maser disk. The inner and outer radius of the disk are about 0.38 and 0.9 mas,
     corresponding to 0.20 and 0.47 pc. The uncertainty on the SMBH mass is $\pm$ 1.5\%. 
     We show the probability density function for the dynamical center position, 
     the SMBH mass and the recession velocity in Fig.~\ref{fig:mcmcpdf}.

		\subsubsection{NGC 5495}
		
    The VLBI map of NGC 5495 shown in Fig.~\ref{fig:VLBImap} uses a
    3$\sigma$ flux cutoff of 6.3 mJy beam$^{-1}$ channel$^{-1}$ with a channel width
    of 125 kHz, in contrast to the 5$\sigma$ flux cutoff we normally used here. This is because there 
     will be no blue-shifted features left when we apply a 5$\sigma$ cutoff and thus we cannot constrain 
     the SMBH mass accordingly. So here we lowered our criteria to have the fitting run successfully. 
     The VLBI image shows a north-south orientation of the maser disk, with
    a P.A. of 176$^{\circ}$ $\pm$ 5$^{\circ}$. We estimate the inclination angle of the disk to be 95$^{\circ}$ $\pm$ 1$^{\circ}$.
    The inner and outer radii of the disk are about 0.22 and 0.65 mas,
    corresponding to 0.10 and 0.30 pc. The high-velocity features show evidence for strong
    warping in the P.A. direction on the red-shifted side. Features with intensity above 
    5$\sigma$ are shown in filled circles on the VLBI map, while those in between 3$\sigma$ 
    and 5$\sigma$ are shown in open circles.

    Our best-fit (with reduced $\chi^{2}$ of 0.22) SMBH mass for NGC 5495 is (1.05 $\pm$ 0.06)
    $\times~10^{7}M_\odot$. The fitted recession velocity V$^{\rm{LSR, opt}}_{0}$ is 6839.6 $\pm$ 67.0 km s$^{-1}$. 
    The uncertainty on the SMBH mass is $\pm$ 5.8\%. The dynamical center is located - 0.051 mas
     from the original reference position (with velocity of 6789.03 km s$^{-1}$)
     on the major axis of the maser disk. Because we have a small number of data points, the reduced $\chi^{2}$  is not a 
     strong indicator of the goodness of fit, and in fact we get value significantly less than unity. 
     We suspect this is caused by the combination of both too few data points and overestimation of the VLBI positional uncertainties.
     As shown on the P-V diagram, the rotation curve is sparsely sampled by only a few data points.
     Overestimation of the positional uncertainties is indicated on the P-V diagram where the two rightmost red-shifted features 
     (close in velocity) are separated much less than their uncertainties in radius, horizontally.   
     The VLBI positional uncertainties are estimated by the empirical relation of 0.5$\Theta/(SNR)$, 
     where $\Theta$ is the projected synthesized beam size and SNR is 
     the signal-to-noise ratio. Due to the low declination of NGC 5495 and the north-south 
     orientation of the maser disk, NGC 5495 has the largest projected synthesized beam among 
     the 5 galaxies in this study. So the positional uncertainty calculated empirically might not correctly reflect the real uncertainty here.  
     We show the probability density function
     for the dynamical center position, the SMBH mass and the recession velocity in Fig.~\ref{fig:mcmcpdf}.

		\subsubsection{MRK 1029}
		
    The GBT spectrum of Mrk 1029 (Fig.~\ref{fig:GBTspec}) does not show the triple-peaked
    profile characteristic of disk systems such as J0437+2456,
    ESO 558$-$G009 and NGC 5495. Instead, there are only two prominent groups of spectral lines centered
    around 8980 km s$^{-1}$ and 9136 km s$^{-1}$. Two recession velocities
    are reported from NED for Mrk 1029 from optical measurements, 9076 $\pm$ 32 km s$^{-1}$ 
    (Huchra et al. 1999) and 9160 $\pm$ 61 km s$^{-1}$
    (de Vaucouleurs et al. 1991). The strongest maser line is at 9136.3 km s$^{-1}$. 
    Our GBT spectrum also shows possible features located near 9550 km s$^{-1}$ and 
    10100 km s$^{-1}$, but our VLBI observation does not cover those velocities. 
        
\begin{figure}
\epsscale{0.75}
\plotone{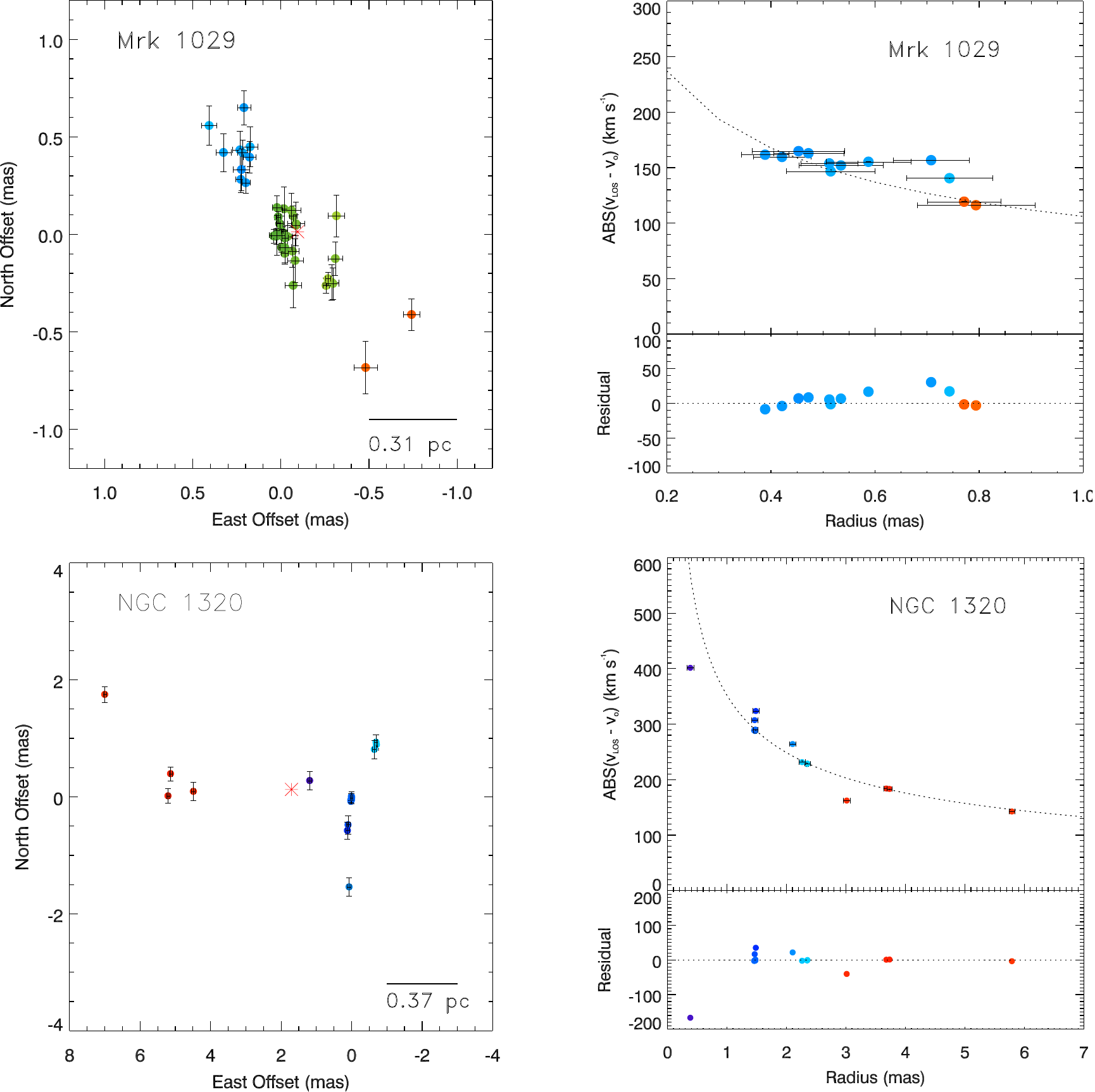}
\caption{The VLBI maps and P-V diagrams for Mrk 1029 and NGC 1320,
  which are not typical Keplerian disk maser systems but do show a
  configuration consistent with circumnuclear rotation. 
  Residuals are shown on the bottom of the P-V diagram. All features shown
  here are above 5$\sigma$.   
\label{fig:VLBImap2}
        }
\end{figure}

\begin{figure}
\epsscale{0.65}
\centering
\plotone{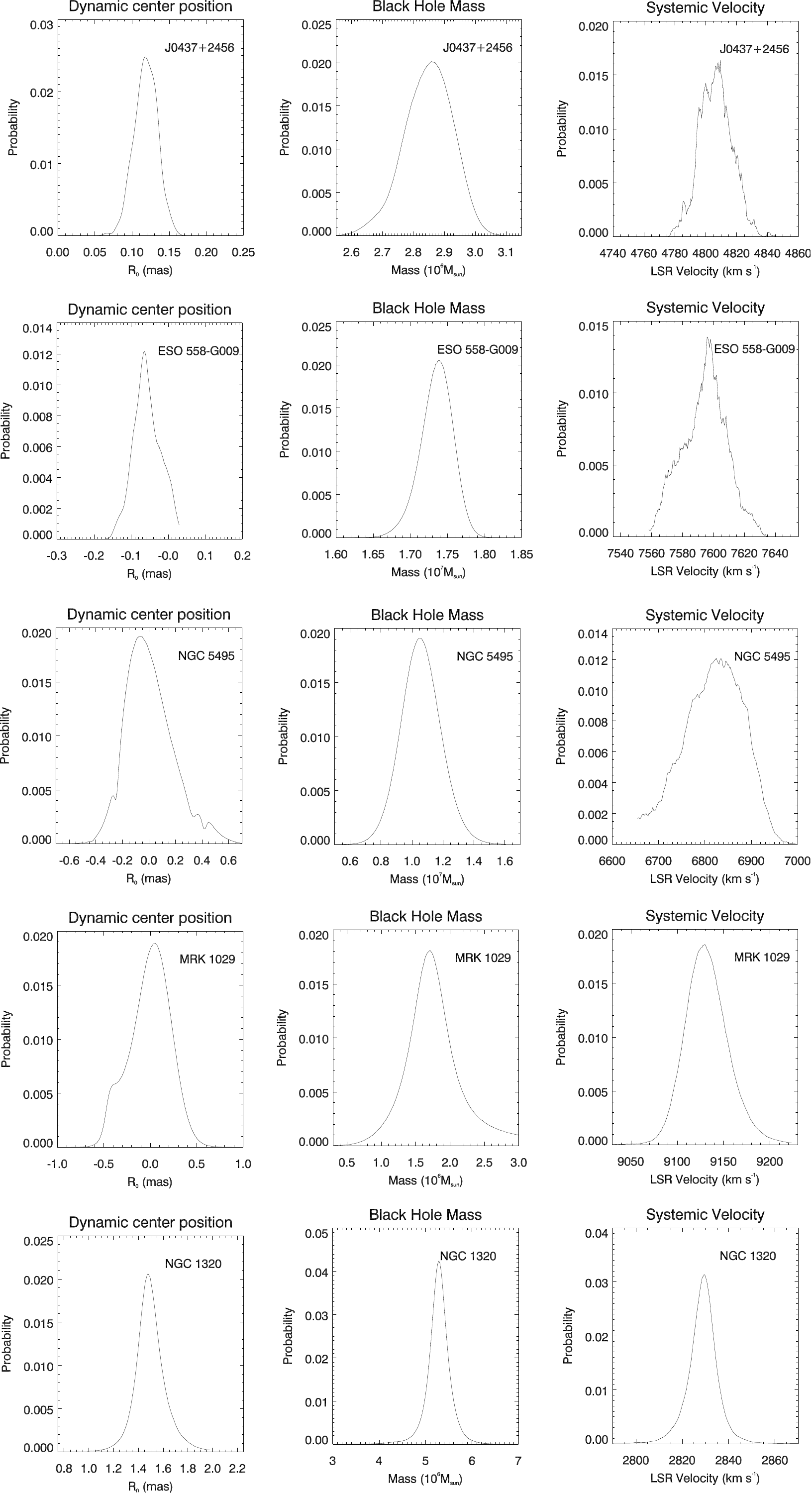}
\caption{Probability density distribution for the fitted dynamical center position, black hole mass
and recession velocity from the Bayesian fitting for the 5 galaxies in this study.
The reduced $\chi^{2}$ for the 5 galaxies are 1.07, 1.28, 0.22, 1.08 and 1.68, respectively. The
small reduced $\chi^{2}$ for NGC 5495 is likely due to small number of data points and 
an overestimation of the VLBI positional 
uncertainty. Please refer to the text for more details.
 \label{fig:mcmcpdf}
        }
\end{figure}

    The colors in the VLBI map of Mrk 1029 in Fig.~\ref{fig:VLBImap2} identify maser
    features between 9100 km s$^{-1}$ and 9220 km s$^{-1}$ as the systemic masers and those
    between 9220 km s$^{-1}$ and 9300 km s$^{-1}$ as the red-shifted masers. Here we set a 5$\sigma$ flux
    cutoff of 7 mJy beam$^{-1}$ channel$^{-1}$ with a channel width of 125 kHz.
    Masers are distributed linearly from the north-east to the south-west
    at a P.A. of 218$^{\circ}$ $\pm$ 10$^{\circ}$, and the red-shifted and blue-shifted
    features are symmetric and roughly equidistant from the systemic features.
    We estimate the inclination angle of the disk to be 79$^{\circ}$ $\pm$ 2$^{\circ}$.
    
    To cope with the ambiguity in identifying the recession velocity, we averaged the two 
    reported velocities above and use 9118 $\pm$ 50 km s$^{-1}$ as a Gaussian prior
    in the Bayesian fitting.  We also set a Gaussian prior of 0.0 $\pm$ 
    0.2 mas for the position of the dynamic center. Due to the ambiguity of maser origin in this galaxy,
    as some of them may not originate in the modeled disk, we used an ``outlier-tolerant'' method rather than Gaussian 
    probabilities when calculating the probability of each model given the data (Reid et al. 2014). 
    This would minimize the effects of deviated data in constraining the model, which is essential 
    for us to reach a self-consistent result here.
        
    We show the best-fit rotation curve and residuals in Fig.~\ref{fig:VLBImap2}, in which all the high-velocity
    features trace the Keplerian rotation well. The best-fit (with reduced $\chi^{2}$ of 1.08) 
    SMBH mass for MRK 1029 is (1.7 $\pm$ 0.45) $\times~10^{6}M_\odot$. 
    The fitted recession velocity V$^{\rm{LSR, opt}}_{0}$ is 9129.9 $\pm$ 26.0 km s$^{-1}$. 
    The uncertainty on the SMBH mass is $\pm$ 27.5\%. The dynamical center is located  0.05 mas
     from the original reference position (with velocity of 9136.86 km s$^{-1}$)
     on the major axis of the maser disk. We show the probability density functions
     for the dynamical center position, the SMBH mass and the recession velocity in 
     Fig.~\ref{fig:mcmcpdf}.

                
Future HI observations of Mrk 1029 could better constrain the recession velocity and thereby aid in
identifying the systemic masers.  Also, the disk interpretation above could be verified if maser features 
redward of 9500 km s$^{-1}$ could be mapped with future VLBI observations.

		\subsubsection{NGC 1320}
    
As in Mrk 1029, the masers in NGC 1320 do not fall into the three distinct velocity groups
that characterize a typical disk maser. In addition to imaging the maser distribution with the 
original channel spacing of 0.86 km s$^{-1}$ (62.5 kHz),
we also averaged the data over several velocity ranges to image faint, broad features.
Fig.~\ref{fig:VLBImap2} shows a map of features detected with SNR greater than 5. 
The maser distribution shows an east-west orientation,
with a P.A. of 75$^{\circ}$ $\pm$ 10$^{\circ}$. No systemic features (near the recession velocity of 
2649 km~s$^{-1}$) were detected in our VLBI observations,
so we cannot estimate the inclination angle using the method described above.
We note, though, that a strong maser feature appeared at the velocity of 2663.0 km~s$^{-1}$ in a 
later (2012 December 19) GBT spectrum. Several maser spots on both the
blue and red-shifted sides are offset from the major axis of the VLBI map, 
suggesting a strong warp or origin of masers different from disk. 
Nonetheless, the map is suggestive of a disk. 
The masers in NGC 1320 cover an angular size of $\approx$ 4 mas, equivalent
to about 1.6 pc at a distance of 38 Mpc. This linear size is on the high end of the distribution
of other maser disk sizes (as listed in Section 4.2).
    
Despite the departures from a characteristic maser disk, we fit the VLBI data of NGC 1320 with the same
``error-tolerant'' Bayesian fitting as we did for Mrk 1029. Fig.~\ref{fig:VLBImap2} shows the best-fit rotation
curve, in which most of the high-velocity features trace the Keplerian rotation well. 
The best-fit (with reduced $\chi^{2}$ of 1.68) black hole mass is (5.3 $\pm$ 0.4) 
$\times~10^{6}M_\odot$ with recession velocity V$^{\rm{LSR, opt}}_{0}$ of 2829.2 $\pm$ 11.4 km s$^{-1}$.
The dynamical center is located 1.5 mas from the original reference position (with velocity of 2545 km s$^{-1}$) 
on the major axis of the maser disk. We show the probability density function for the dynamical center position, 
the SMBH mass and the recession velocity in Fig.~\ref{fig:mcmcpdf}. As the maser profile is highly variable,
future VLBI observations could potentially catch more systemic masers and high-velocity masers to enable a better 
characterization of the maser system in NGC 1320 (e.g. test for a warped disk). 

	\subsection{Possible sources of systematic error and overall SMBH masses}
	
In addition to the formal fitting errors we estimated above, there are several other factors that
contribute to the uncertainty in estimating the black hole mass. The SMBH mass depends directly
 on the galaxy distance, which we estimate from the Hubble law.
So any departure from the pure Hubble flow will affect our distances and SMBH masses. 
A typical peculiar velocity of 300 km~s$^{-1}$ (e.g. Masters et al. 2006) would cause an uncertainty 
on the SMBH mass of 6.3\% for J0437+2456, 3.9\% for ESO 558$-$G009, and 4.4\% for NGC 5495, 
to be added in quadrature.

In our fitting model, we assume that all high-velocity features lie on the mid-line of the disk.
To account for departures from the mid-line, a velocity correction is needed when converting
rotation velocity to the LOS velocity. Depending on the mass of the SMBH, the radius of the maser's orbit, 
and the angular departure from the mid-line, one may be able to detect line-of-sight accelerations even
for the high-velocity maser lines through spectral monitoring.  For a high-velocity feature located 
away from the mid-line by an angle of $\phi$, the LOS velocity would deviate from the Keplerian
rotation velocity by about 1-$cos(\phi)$.  Since the deviation of maser features from the
mid-line are more likely to be distributed between zero and the max angle $\phi_{max}$, the mean
deviation on the rotation velocity will be smaller than 1-$cos(\phi_{max})$. In the best
measured disk maser galaxy NGC 4258, most of the high-velocity features
lie within 5$^{\circ}$ from the midline, and the maximum deviation from the midline is 13$^{\circ}$ as shown 
by few maser features (Humphreys et al. 2008). If not corrected, a 13$^{\circ}$ deviation would introduce 
a 2.6\% adjustment on the rotation velocity and a 5.2\% under-estimation
 on the SMBH mass. (The uncertainty of the fitted black hole mass ${\Delta M}$ depends on the
 deviation of the velocity of the high-velocity features from the best-fit rotation curve ${\Delta V}$
 as $\frac{\Delta M}{M} = (1+\frac{\Delta V}{V})^{2}-1$.)
    
In this paper we estimated the inclination angle $i$ prior to the disk fitting, and then fixed it 
during the fitting to speed convergence.
The effect of fitting the inclination angle is similar to the deviation of the high-velocity
features from the mid-line, as a sin $i$ factor is applied to the rotation velocity.
The difference is that the inclination angle is a systematic effect on all rotation velocities and
it causes an under-estimate of the SMBH mass. The inclination angle of the maser disks currently 
known are all close to perfectly edge-on ($i$ = 90$^{\circ}$) with a typical deviation of 10$^{\circ}$ 
(in the best measured case of NGC 4258, the inclination angle at the radius of the systemic masers 
is about 81.7$^{\circ}$ (Humphreys et al. 2013)). 
In our study, the typical deviation from edge-on is 10$^{\circ}$, which corresponds to a 
change of rotation velocities by 1.5\% and the fitted black hole mass by 3\% from a perfectly edge-on system.

As mentioned in Section 2.2, the absolute positional uncertainty of the reference feature in ``self-calibration" observations may
induce a systematic positional uncertainty for each maser spot. This positional uncertainty is given by
$(\Delta\nu/\nu_{i})\theta_{ref}$ (Thompson et al. 2001). Here the frequency for maser feature $i$ is $\nu_{i}$,
where $\Delta\nu$ is the frequency difference between the reference feature and feature $i$, and $\theta_{ref}$ is the positional uncertainty 
of the reference feature. The resulting positional uncertainties should be added in quadrature to the formal fitting uncertainty for
each maser spot. For ESO 558$-$G009, the positional uncertainty of the reference feature is 72 mas. 
Considering the typical frequency difference of 40 MHz, this gives a positional uncertainty
of 0.13 mas. We have added this uncertainty in quadrature to the formal fitting uncertainty for
each maser spot, and the fitted SMBH mass remains 1.70 $\times~10^{7}M_\odot$, with the uncertainty increased to 2.3\%.
For NGC 5495, the positional uncertainty of the reference feature is about  300 mas, with a
typical frequency difference of 30 MHz, the additional positional uncertainty is about 0.4 mas.
These yield a SMBH mass of 1.04 $\times~10^{7}M_\odot$, with the uncertainty increased to 17\%.
Because of the uncertain nature of the maser structures in NGC 1320 and Mrk 1029, 
we didn't include the position uncertainty in the estimates of their SMBH masses. 

Finally, adding all the systematic uncertainties mentioned above, the SMBH mass results are
(2.9 $\pm$ 0.3) $\times~10^{6}M_\odot$ (9.2\%) for J0437+2456,
(1.7 $\pm$ 0.1) $\times~10^{7}M_\odot$ (7.5\%) for ESO 558$-$G009,
and (1.1 $\pm$ 0.2) $\times~10^{7}M_\odot$ (18.6\%) for NGC 5495.

\section{DISCUSSION}

	\subsection{The ``clean'' disk dataset}
Here we compile a sub-sample of VLBI-confirmed, ``clean'' disk maser galaxies from the
disk maser sample listed in Pesce et al. (2015), and we provide the SMBH mass and the inner and
outer radius of the maser disk for each galaxy, as listed in Table~\ref{table:diskdata}.
We define a clean disk as requiring that 1) the P-V diagram derived from VLBI observations is
consistent with Keplerian rotation and 2) all the maser components come from the disk,
with no jet or outflow contamination. Based on these criteria, NGC 1068, Circinus, NGC 4945,
NGC 1386 and NGC 3079 are not included in our list. The purpose of the list is not to be
as complete as possible, but to identify a physically homogenous sample. Our final
list contains 15 sources, including the seven from Kuo et al. (2011), three from this work
(J0437+2456, ESO 558$-$G009, and NGC 5495), NGC 4258 (Humphreys et al. 2013),
NGC 3393 (Kondratko et al. 2008), NGC 5765b (Gao et al. 2016), IC 2560 (Wagner et al. in prep)
and UGC 6093 (Zhao et al. in prep).

We note that uncertainties on the 
SMBH mass are not calculated in the same way for the whole ``clean'' disk sample, in which sources 
from Kuo et al. (2011) are calculated as this work, 
NGC 4258 and NGC 5765b are analyzed using the 3-D modeling of the disk, while the 
uncertainties reported for NGC 3393, IC 2560 and UGC 6093 are just formal fitting errors. So we
calculated the systematics mentioned in Section 3.3 for NGC 3393, IC 2560 and UGC 6093 and 
add them in 
quadrature to their formal fitting error, except for the uncertainty caused by absolute position 
uncertainty of the reference maser feature. The SMBH mass uncertainties listed in 
Table~\ref{table:diskdata} contain these systematics.

\subsection{Estimation of the maser disk radii from both single-dish spectra and VLBI maps}

Our use of the single-dish spectra is complementary to the VLBI data, in the sense that even though single-dish spectra do not provide spatially-resolved constraints of the position of maser spots, the sensitivity level reached in stacked single-dish observations is usually higher than VLBI due to more available observations. For our study, all 15 sources listed in the ``clean" disk sample have multiple single-dish observations with the GBT (with a typical 1$\sigma$ noise level of less than 1 mJy per 24.4 kHz after combining all the data), while usually the imaging sensitivity reached in VLBI observations is between 5 and 10 mJy per beam per 125 kHz. So there might be more extended emission revealed in the single-dish spectra, that is not picked up by our VLBI observations.

We used single-dish spectra from Pesce et al. (2015), which has combined
 all available GBT spectra for each of the 15 sources, thus providing unprecedented sensitivity. 
 We estimate the maser disk radii with $\rm {r_{edge} = \frac{GM}{(V_{edge}-V_{0})^{2}}}$, 
 where M is the measured SMBH mass, V$_{0}$ is the best-fit recession velocity of the SMBH from VLBI data, 
 V$_{edge}$ are the velocity boundaries (identified with SNR of maser feature's intensity greater than 5), corresponding to the minimum and maximum Keplerian rotation velocity on both the red-shifted and blue-shifted side. From this we get the inner and outer radii of both the red-shifted and blue-shifted masers, and we select the minimum and maximum radii for the high-velocity masers. We estimate uncertainties in maser disk radii from both the SMBH mass measurement and the fitted recession velocity. Here we added a 5.5\% uncertainty on the SMBH mass in quadrature to the formal fitting error to account for the uncertainty on the galaxy distance \footnote{We consider two recent results on $\Ho$, which are 73.00 $\pm$ 1.75 km~s$^{-1}$ Mpc$^{-1}$ (Riess et al. 2016) and 67.8 $\pm$ 0.9 km~s$^{-1}$ Mpc$^{-1}$ (Planck Collaboration et al. 2015) to arrive at the conservative estimate of 5.5\% uncertainty on $\Ho$.}. We note that this radius estimation is based on the assumptions that all masers we included trace the Keplerian rotating disk and they are also located on the mid-line of the edge-on disk. These assumptions could be violated by maser flares or masers generated from outflows. 
 
From our VLBI maps, we estimate the range of radii for the systemic masers and high-velocity masers separately. 
For the high-velocity masers, we measure the projected separation between the maser positions and the fitted dynamical center, directly. 
Uncertainties on the maser radii come from position uncertainties on the maser spots themselves and the uncertainties on the distances 
to the galaxies. For the distance calculation, we also apply a 5.5 \% uncertainty on $\Ho$, as mentioned above. For the systemic masers, 
there are two ways to estimate the radius. First,
if the accelerations (velocity drift with time) of the systemic features have been measured from spectral monitoring,
we apply that together with the measured SMBH mass to constrain the radius
as $r = \sqrt{\frac{GM}{a}}$. Second, the radius can be estimated from the slope of the systemic
features on the P-V diagram, $\Omega = \sqrt{\frac{GM}{r^{3}}}$. 
The second method does not work well if the systemic features are distributed over a large range
of radii, as different slopes on the P-V diagram will blend together. So here we use the first method when applicable.  

Finally we compare the maser disk radii estimated from both single-disk spectrum and the VLBI maps. If the two results are consistent within 3$\sigma$, we use the radius derived from VLBI as the final value. Otherwise we still apply the VLBI result as the final value, but we add the difference between the two methods in the uncertainties. We note that the inner radii estimated from the two methods are consistent in all other galaxies except NGC 1194. And NGC 3393 remains the only case where the inner radius estimated from the systemic masers is significantly smaller than that estimated from high-velocity VLBI map, as originally reported in Kondratko et al. (2008). For the outer radius, inconsistency is seen in NGC 4388, NGC 2273, NGC 1194, NGC 3393, UGC 3789, NGC 2960 and UGC 6093. In the first 4 of these galaxies, the single-dish spectrum shows a continuous maser distribution between the high-velocity features and systemic features (see Fig.1 in Pesce et al. 2015), making it difficult to assign the velocity range associated with the maser disk. We list the inner and outer radii for each maser disk in Table~\ref{table:diskdata}. Radii which show inconsistency between the single-dish spectrum estimation and VLBI results are marked by an asterisk. The original measured radii for each galaxy from both methods are listed in Table~\ref{table:radiidata}.

	\subsection{SMBH mass versus maser disk size}
	
Disk masers provide the best measured extragalactic SMBH masses
(e.g. Kormendy \& Ho 2013; Miyoshi et al. 1995; Kuo et al. 2011; this work). However,
the low detection rate even in type 2 AGN ($<$ 1\%, Braatz et al. in prep)
prevents accumulating a large sample. To improve the efficiency of future surveys, several 
studies have tried to correlate the incidence of maser emission with host galaxy
properties, either in the optical (Zhu et al. 2011; Constantin 2012; van den Bosch et al. 2016), 
radio (Zhang et al. 2012) or X-ray (Zhang et al. 2006, 2010), but no clear conclusion has been made.
A better understanding of the mechanism that generates disk masers will also provide 
important guidance for future disk maser surveys. Below we concentrate on the mechanisms
 that determine the inner and outer radius of the maser disk.

Theoretical models have been put forth to explain the generation of maser emission around
a SMBH. Neufeld et al. (1994) proposed that X-ray emission from the central AGN
provides the heating that generates a suitable environment for population inversion and maser emission.
In their model, X-rays irradiate the slightly
tilted and warped molecular disk, which is formed on the outer part of the accretion disk.
As the molecular layer is heated, the abundance of H$_{2}$O molecules increases.
With the molecular hydrogen density between 10$^{7}$-10$^{11}$ cm$^{-3}$, gas temperature
between 400-1000 K and water abundance x(H$_{2}$O) = n(H$_{2}$O)/n(H$_{2}$) $\gtrsim$ 10$^{-4}$,
luminous maser emission can be generated. Based on standard accretion disk theory
(Shakura \& Sunyaev 1973), Neufeld \& Maloney (1995, hereafter NM95) suggested that the outer radius 
for masing is determined by a phase transition from molecular to atomic gas, while the inner
radius is caused by the flattening of the disk, making it no longer exposed to
the central X-ray source. So the gas there would be too cold to pump masers.
Collison \& Watson (1995) include dust grains in their calculation of maser emissivity and
show that cold dust (with respect to warm molecular gas) is essential for maintaining maser emission.
Based on these studies, the AGN's luminosity, the inclination and warp of the disk, and the 
gas and dust content of the molecular disk together determine the radii at which maser emission forms.

Wardle \& Yusef-Zadeh (2012, hereafter WY12) reported an upper envelope to the disk maser
outer radii, which scales with SMBH mass as $R_{out} \approx 0.3 M_{7}$ pc
(with $M = M_{7}\times10^{7}M_\odot$), based mainly on the results of Kuo et al. (2011).
To better explore the relation between SMBH mass and maser disk size, here we fit the 
inner and outer radius of the maser disk with the SMBH mass in the the form of 
$Log (R) = \alpha Log (M) + \beta \pm \epsilon$. We note that the uncertainties 
on the measured inner and outer radius of NGC 4258 are about an order of magnitude smaller than in 
other sources, making NGC 4258 dominate any linear fitting result and possibly suppressing any 
general trend between maser disk size and SMBH mass.
Nevertheless, we calculate the Spearman's rank correlation coefficient $\rho_{S}$ for the inner and outer radius case, giving
 0.71 and 0.62, respectively. Their two-tail probability are 0.01 in both cases. Since these coefficients do suggest the existence 
of some correlation, here as a preliminary check, we fit the data with equal uncertainties (uniform weighting)
on the maser disk radius between different sources and get:      

   $$ \rm{Log(R_{in}/pc) = (0.60 \pm 0.17) \times Log(M_{SMBH}/M_\odot) - (5.21 \pm 1.17) \pm 0.22}$$
   $$ \rm{Log(R_{out}/pc) = (0.57 \pm 0.16) \times Log(M_{SMBH}/M_\odot) - (4.50 \pm 1.17) \pm 0.22}$$

\begin{figure}
\epsscale{1.0}
\plotone{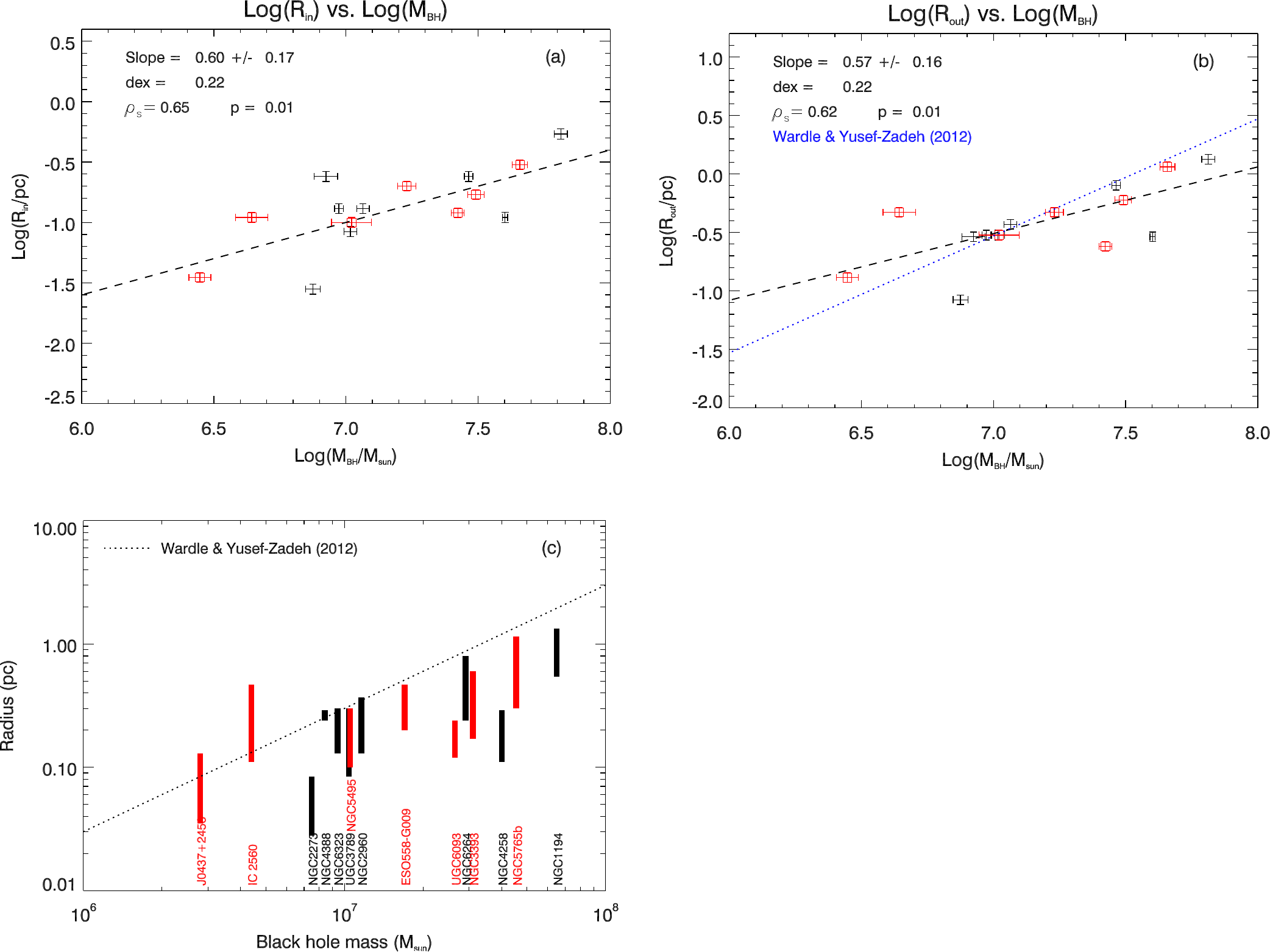} 
\caption{Relation between the maser disk radius and the central SMBH mass. Maser disks
 reported in Kuo et al. (2011) are shown in black, while new results are shown in red. We
 show the inner and outer radius of maser disks versus SMBH mass in plots (a) and (b), with
 the best-fit result with uniform weighting shown as the black dashed line, while the blue dotted line in plot (b) shows
 the scaling relation reported in Wardle \& Yusef-Zadeh (2012). We also show the maser
 disk extension (from the inner to the outer radii, as shown by the vertical bars) versus
 SMBH mass, following the plot style in Wardle \& Yusef-Zadeh (2012), in plot (c). The dotted
 line also shows the scaling relation of the outer radius of maser disks vs. SMBH mass as
 reported in Wardle \& Yusef-Zadeh (2012). 
 \label{fig:mvsr}
        }
\end{figure}

We show the fitting results in Fig.~\ref{fig:mvsr} (a) and (b) respectively, in which the black dashed
 line shows the best-fit result in both cases, while the blue dotted line in Fig.~\ref{fig:mvsr} (b) shows
 the relation from WY12. For direct comparison to WY12, we show the extent of the maser disks by 
 the vertical bars in Fig.~\ref{fig:mvsr} (c). Targets from Kuo et al. (2011) are shown in black, while 
 others are shown in red. The black dotted line marks the relation from WY12. 
We do not include Sgr A*, which is used in WY12 but does not have disk masers.
   
In general, new results from this work and others strengthen the empirical upper envelope reported 
in WY12 (as shown in Fig.~\ref{fig:mvsr} (c)), although the slope we get for the outer radii
under uniform weighting is not consistent with the slope of 1.0, which was used in WY12. 
And we note that although NM95 derived $R_{out} \propto L_{41}^{-0.426} M_{8}^{0.617}$
(Equation 4 in NM95), which is consistent with our fitting results, their model 
was based on the assumption of steady accretion from the pc-scale maser
disk to the central SMBH, which may not be valid in disk maser galaxies
(e.g. Gammie, Narayan \& Blandford 1999).

The SMBH mass sets an additional constraint on the inner radius of the maser disk.
Maser amplification requires not only spatial alignment of molecular clouds along the gain path, 
but also velocity coherence along the LOS. The typical velocity coherence for the 22 GHz H$_{2}$O
disk maser is about 1 km s$^{-1}$ (the typical linewidth seen in disk masers). 
Velocity coherence path lengths shorten at smaller disk radii, as the rotation velocity increases. 
Gao et al. (2016) show $R_{in} \propto$ M$_{SMBH}$ under such velocity coherence requirement.
However, this differs from the fit determined above.
So we conclude that the inner radius of the maser disk may not be determined solely
 by velocity coherence.

	\subsection{An empirical relation between WISE luminosity and maser disk outer radii}
	
As mentioned above, an AGN luminosity may determine the range of radii over which water masers amplify. 
Below we examine the relation between maser disk size and AGN luminosity. 
[O III] and X-ray 2-10 keV luminosity are often used as indicators for the isotropic AGN luminosity 
(e.g. Zakamska et al. 2003; Heckman et al. 2004; Liu et al. 2009). 
For the 15 galaxies in our ``clean" disk maser sample, only 9 have 2-10 keV X-ray
measurements (Masini et al. 2016; Castangia et al. 2013 and references
therein), and 14 have [O III] measurements (Table~\ref{table:diskdata}).  
Most of our disk maser galaxies are Compton thick and 3 of the 9 X-ray measurements 
have only a lower limit on the intrinsic column density, which affects the estimation
of the intrinsic X-ray luminosity. The [O III] data may suffer from intrinsic reddening
(Greene et al. 2010).  So besides using [O III] and X-ray data, here we explore the use of 
the Wide-Field Infrared Survey Explorer (WISE) Band 3 ($\lambda$ =12 $\mu$m) luminosity
(Wright et al. 2010) as an approximation for the AGN luminosity. 
        
We choose the Band 3 data over other bands (3.4, 4.6 and 22 $\mu$m) for two reasons.  First, there is a significant
contribution from stellar photospheric emission at wavelengths shorter than 10 $\mu$m
(e.g. Fu et al. 2010).  Second, we can estimate the AGN bolometric luminosity
L as $L = \int L_{\nu}d\nu =  \int L_{\lambda}d\lambda \approx \lambda L_{\lambda}$ 
near $\lambda$ =10 $\mu$m when U $\geq 10^{7}$ (Drane \& Li 2007), where U is the
radiation parameter. See the Appendix below for the derivation of U in disk maser galaxies.
    
\begin{figure}
\epsscale{0.8}
\plotone{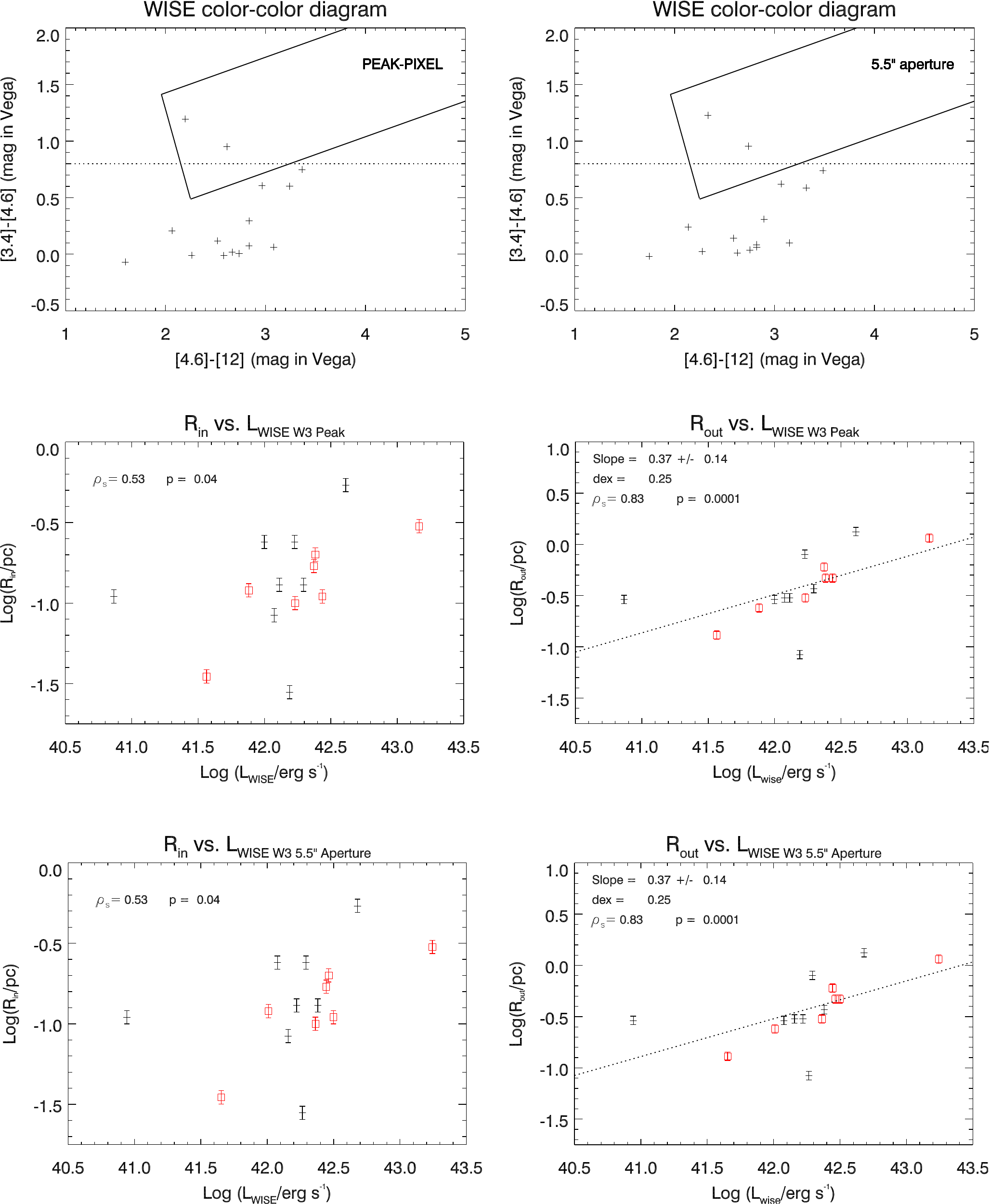} 
\caption{In the upper panel we show the WISE color-color plot (W1-W2 vs. W2-W3)
for our disk maser sample for both the central pixel photometry case and the 5.5" aperture photometry
case to help identify their MIR-radiation origin. For guidance, we plot the color-color wedge
that luminous AGN may fall in from Mateos et al. (2012) in solid lines and the W1-W2
of 0.8 in dotted line (Stern et al. 2012). Only NGC 1194 and NGC 4388 in our sample fall 
within the color-color wedge. Plots on the middle and bottom panel show the
relation between inner and outer radii of maser disks with AGN luminosity measured from
WISE W3 (12 $\mu$m) peak pixel and 5.5" aperture photometry. Maser disks reported
in Kuo et al. (2011) are shown in black cross, while new results are shown in red open square.
The dotted line in the right panels indicate the best-fit results, under uniform weighting. 
On the upper left corner in each subplot, we give the Spearman's rank correlation coefficient
$\rho_{S}$ together with the two-side possibility p-value. The best-fit slope and scatter are also
shown when possible.
 \label{fig:lvsr}
        }
\end{figure}

The spatial resolution of WISE at Band 3 is about 7.4". To eliminate contamination
from star formation regions in the host galaxy, we use the 5.5" aperture
photometry archive data, which has the smallest aperture among the WISE
datasets and best match the Band 3 resolution. 
We also measure the background subtracted flux from the central pixel from the
WISE Band 3 online image. Each pixel covers 1.375" on the image. Galaxies in
our sample have distances between 7.6 Mpc and 153.2 Mpc, so the pixel size ranges from
50 pc to 1 kpc. To test whether the observed IR emission originates from the AGN,
we examine the WISE color-color plot (W1-W2 vs. W2-W3) for our disk maser sample
(Fig.~\ref{fig:lvsr}). For guidance, we plot the color-color ``wedge" that luminous AGN may fall in 
(Mateos et al. 2012) and the empirical mid-infrared criterion for AGN identification,
which is W1-W2 of 0.8 (Stern et al. 2012), in dotted line.
Clearly most of our disk maser galaxies do not fall in the wedge nor do they lie above the
W1-W2 of 0.8 line. This is not surprising as the completeness of the color-color wedge selection
is a strong function of X-ray luminosity, as stated in Mateos et al. (2012).
Below 10$^{43}$ erg s$^{-1}$, where all nine disk maser galaxies with published X-ray
measurements are located (Masini et al. 2016, Castangia et al. 2013),
the WISE color-color wedge only selects less than 20\% of type 2 AGN (Mateos et al. 2012).
The majority of disk maser galaxies in our sample have W2-W3 between 2 and 3, and
W1-W2 near zero. Comparing with the color-color diagram for different types of sources
as shown in Yan et al. (2013), we see our WISE luminosities may have contamination from
star formation and we cannot distinguish that from AGN emission.

Nevertheless, our logic here is that if we can see any correlations between the ``star formation
 contaminated" WISE luminosity and maser disk properties (e.g. inner and outer radius), then 
 such correlation should be strengthened when the intrinsic luminosity from the AGN has been used 
 instead, as star formation activities should not affect the generation of maser emission. So we plot 
 the inner and outer maser disk radii against the WISE Band 3 luminosity measured from both the 
 peak pixel and 5.5" aperture in the middle and bottom panel in Fig.~\ref{fig:lvsr}, respectively.
 
 These plots suggest a correlation between the outer disk radius and
 the WISE luminosity, both in the peak pixel data and the 5.5" aperture data. Here we calculate the 
 Spearman's rank correlation coefficient $\rho_{S}$ as a non-parametric test, giving 0.83 with the 
 two-tail probability of 0.0001 for both cases. Such a high coefficient suggests a positive correlation.
 In contrast, no clear correlation is found between the inner disk radius and the WISE luminosity. 
 The calculated Spearman's rank correlation coefficient is 0.53 with the p-value of 0.04, which disfavor 
 a correlation.
 
 To better characterize the correlation between the outer radius and the WISE luminosity, here we fit 
 the data in the the form of $\rm{Log (R) = \alpha Log (L) + \beta \pm \epsilon}$. Similar to the previous
 section, here we assign a 10\% uncertainty for the disk radius in each galaxy, which gives equal weighting 
 for all galaxies in the fitting. This would avoid NGC 4258 dominant the fitting and yields results with
 unreasonably high confidence levels. The best-fit results using the peak pixel data is:
   $$ \rm{Log(R_{out}/pc) = (0.37 \pm 0.14) \times Log(L_{WISE W3}/(erg~s^{-1})) - (16.19 \pm 6.00) \pm 0.25}$$
   
\noindent The fit from the WISE 5.5" aperture data is consistent with that from the peak pixel data. 
However, both fits are strongly influenced by a small number of data points at the low and high 
ends of the luminosity distribution.
We show the results on the middle and bottom panel of Fig.~\ref{fig:lvsr}.  

\begin{figure}
\epsscale{1.0}
\plotone{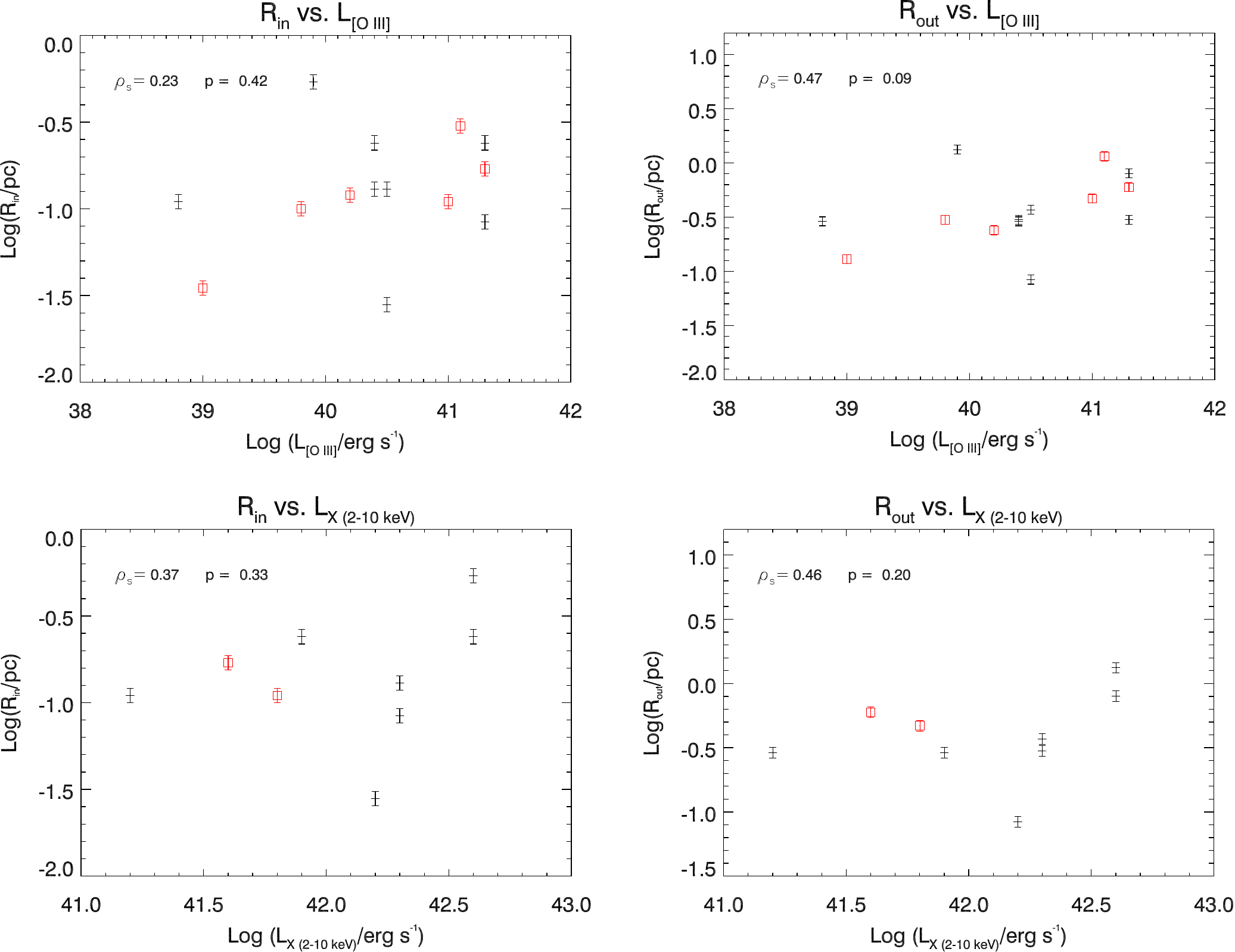} 
\caption{Correlation between inner and outer radii of maser disks with AGN luminosity
measured from [O III] and X-ray 2-10 keV data. Maser disks reported in Kuo et al. (2011)
are shown in black cross, while new results are shown in red open square, when their [O III] or X-ray
measurements are available. 
we give the Spearman's rank correlation coefficient
$\rho_{S}$ together with the two-side possibility p-value.
 \label{fig:lvsr2}
        }
\end{figure}

\begin{figure}
\epsscale{1.0}
\plotone{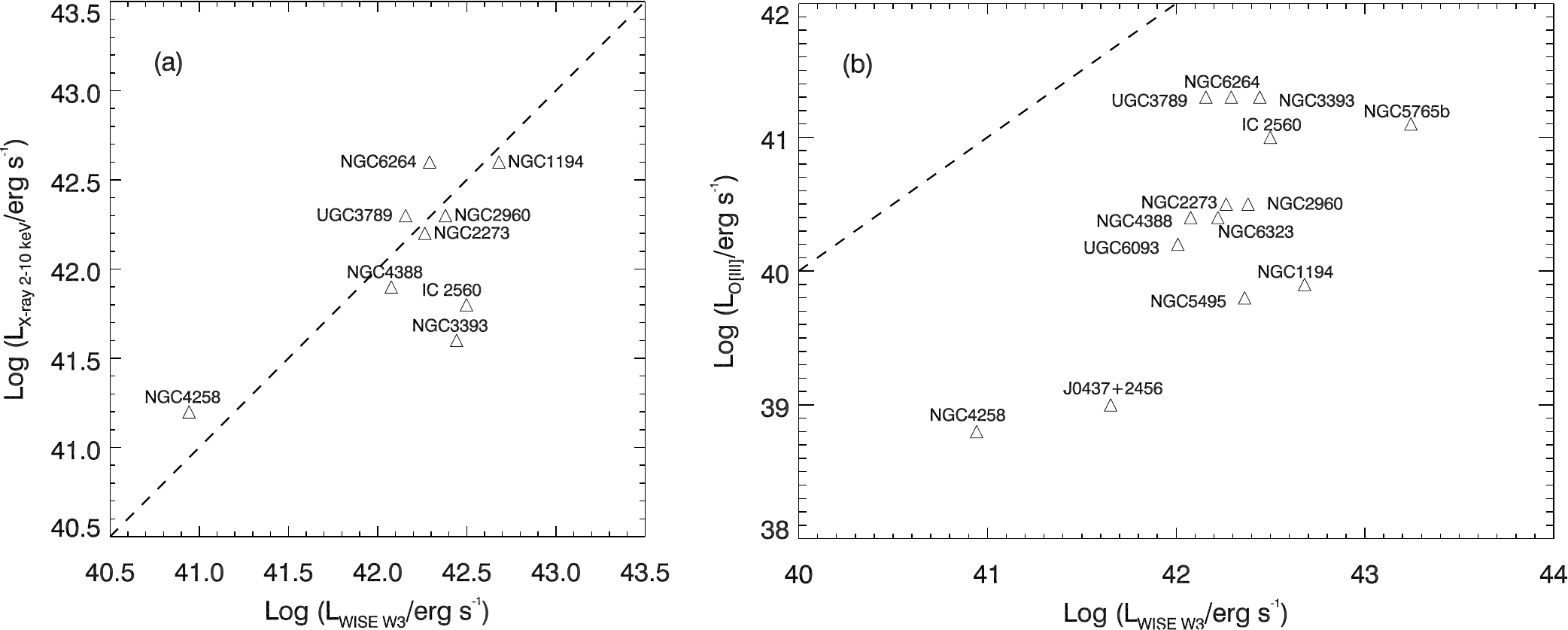} 
\caption{Comparison between the WISE Band 3 luminosity with (a) the X-ray 2-10 keV
luminosity and (b) the [O III] luminosity. The dashed line shows where the two luminosities are equal,
in both plots. The WISE Band 3 luminosities come from our measurement of the central
pixel value, [O III] data come from Greene et al. (2010) and SDSS online listings,
and X-ray 2-10 keV results all come from Castangia et al. (2013). We list the WISE
luminosity and [O III] luminosity in Table~\ref{table:diskdata}.     
 \label{fig:lcomp}
        }
\end{figure}

We also calculate $\rho_{S}$ between the maser disk size and both the absorption-corrected 2-10 keV luminosity
and [O III] luminosity, though they are all less than 0.5, suggesting no evident correlation.
We show these results in Fig.~\ref{fig:lvsr2}.  
To investigate further, here we compare the 2-10 keV luminosity and [O III] luminosity against 
WISE luminosity in Fig.~\ref{fig:lcomp}. In general, there is a good consistency between the 
X-ray data and WISE data, though the small number of available X-ray measurements 
prevents a more systematic comparison. As for the [O III]  data, the overall offset and scatter 
may reflect the uncorrected reddening, as only Galactic absorption has been corrected in 
the [O III] data here.
   
Masini et al. (2016) also compared the inner and outer radii of maser disk in a sample of 14 maser galaxies 
to the X-ray 2-10 keV luminosity measured from the Nuclear Spectroscopic Telescope Array (NuSTAR), 
and they didn't find a correlation neither. Though we note that several ``non-clean'' 
disk masers have been included in their study (e.g. NGC 1068, NGC 1386, NGC 3079, 
NGC 4945 and Circinus). Besides, the X-ray luminosities we used here are measured by either
XMM or Chandra at the wavelength roughly between 0.1 and 10 keV, whereas NuSTAR covers 
the higher energy band between 3 to 79 keV, which is less sensitive to absorption and thus
not good coping with absorption corrections. Nevertheless, our current result and results from 
Masini et al. (2016) do not suggest the existence of correlation between maser disk size and X-ray
luminosity. Considering the good consistency between the X-ray data and WISE data we mentioned above,
a larger sample of disk masers with X-ray measurements would be better to verify the correlation between
maser disk size and AGN luminosity here.

One way to interpret the R$_{out}$-L$_{WISE}$ correlation is that since 
we see a tentative correlation between maser disk size and SMBH mass 
(as in previous section), and it is well known that SMBH mass correlates 
with galaxy bulge mass (e.g. Kormendy \& Ho 2013), which can convert to bulge 
luminosity. So we might expect a correlation between maser disk size and 
mid-infrared luminosity, if it contains luminosity from both the AGN and the 
galaxy bulge. However, this interpretation might not hold true, since we also 
see a correlation between R$_{in}$ and SMBH mass but we don't see a strong
correlation between R$_{in}$ and L$_{WISE}$ here.

As we've derived in the appendix, if the outer radius of the maser disk is constrained by the molecular
gas kinematic temperature as it approaches $T_{min} \approx$ 400 K, the slope derived between R$_{out}$ 
and mid-infrared luminosity should be 0.5, with an intercept of 22.18. Our current best-fit result
is consistent with such a slope and intercept, but a larger sample of clean disk maser galaxies are needed to 
verify the result. In addition, higher-resolution than the 7.4" for WISE  band 3 are needed to eliminate 
contamination from star formation and stars in the host galaxy and verify the correlation.


\section{CONCLUSIONS}

We present high-resolution VLBI observations of five new maser
systems: J0437+2456, ESO 558$-$G009, NGC 5495, Mrk 1029 and NGC 1320.
The maser systems in J0437+2456, ESO 558$-$G009 and NGC 5495 clearly delineate
edge-on maser disks and we measure their central black hole masses by
fitting a Keplerian curve to their rotation profiles. The best-fit SMBH mass results are
(2.9 $\pm$ 0.3) $\times~10^{6}M_\odot$ for J0437+2456,
(1.7 $\pm$ 0.1) $\times~10^{7}M_\odot$ for ESO 558$-$G009,
and (1.1 $\pm$ 0.2) $\times~10^{7}M_\odot$ for NGC 5495. 

The maser spectrum in Mrk 1029 does not show a typical disk maser configuration, 
though the VLBI data could be fitted by the disk model, from which we get the SMBH 
mass of (1.64 $\pm$ 0.45) $\times~10^{6}M_\odot$. Future VLBI observations that 
detect maser features at higher rotation velocities could confirm the disk model.  
For NGC 1320, the maser distribution presents a more complicated
picture. Some of the maser spots may trace outflowing gas while others may
come from a disk.  Our data do not provide a robust black hole mass in this case.
The preliminary SMBH mass under the Keplerian disk model is 
(5.3 $\pm$ 0.4) $\times~10^{6}M_\odot$.
 
With these new measurements, we define a new sample of VLBI-confirmed ``clean'' maser
disks and we use this sample to investigate relationships
between maser disk size, black hole mass, and AGN luminosity.
Relating maser disk radii to the SMBH mass, our new measurements together with
other works confirm the empirical upper envelope of $R_{out} \propto 0.3 M_{SMBH}$ 
reported in Wardle \& Yusef-Zadeh (2012). We use currently available X-ray 2-10 keV
measurements, [O III] measurements, and WISE 10 $\mu$m measurements as
indicators for the AGN luminosity for the disk maser sample and
identify a possible correlation between maser disk outer radii and WISE
luminosity with a high Spearman's rank coefficient of 0.8.  We do
not see similar correlations with X-ray 2-10 keV luminosity or [O III]
luminosity data. If confirmed, this correlation suggests that the outer
radius of maser disk is determined by the molecular gas kinetic
temperature approaching $T_{min} \approx$ 400 K, which is essential
for pumping water masers.

\vskip 0.5truecm
F.G. gratefully acknowledges support provided by
the National Radio Astronomy Observatory (NRAO) through
the Grote Reber Doctoral Fellowship, and support in part by the
Major Program of National Natural Science Foundation of
China (grants 11590780 and 11590784) and the Strategic
Priority Research Program “The Emergence of Cosmological
Structures” of the Chinese Academy of Sciences, grant No.
XDB09000000. The National Radio Astronomy Observatory is
a facility of the National Science Foundation operated under
cooperative agreement by Associated Universities, Inc. This
research has made use of the NASA/IPAC Extragalactic
Database (NED), which is operated by the Jet Propulsion
Laboratory, California Institute of Technology, under contract
with the National Aeronautics and Space Administration.

\vskip 0.5truecm
\noindent
{\it Facilities:} \facility{VLBA, VLA, GBT, Effelsberg}

\appendix

\section{L(R$_{out}$) for water masers} 
\label{a:app}

If the outer radius R$_{out}$ of a circumnuclear water maser disk is set by the requirement
that the kinetic temperature is at least T$_{min}$, then the inverse-square law suggests it
varies with AGN luminosity L as R$_{out} \propto L^{1/2}$ in radiative equilibrium. The upper
energy level of the 6$_{16}$ - 5$_{23}$ 22 GHz water transition, expressed as a temperature,
is E$_{u}$/k $\approx$ 640 K, so 22 GHz water masers must originate in molecules with kinetic
temperatures approaching 640 K. The minimum kinetic temperature is
T$_{min}$ $\approx$ 400 K (Neufeld et al. 1994). 

The water molecules are mixed with dust, and the dust temperature T$_{d}$ is determined
by the local radiation energy density u$_{rad}$. From Equation 24.19 and 24.20 in Draine (2011),

\begin{equation}
T_{d} \approx 20~U^{1/6} \label{eqn:eqA1}
\end{equation}

where the radiation parameter U is defined by

\begin{equation}
U \equiv \frac{u_{rad}}{u_{\star}} \label{eqn:eqA2}
\end{equation}

and $u_{\star} \approx 1.05 \times 10^{-12}$ erg cm$^{-3}$ is the interstellar radiation energy
density in the Solar neighborhood. Thus in CGS units

\begin{equation}
T_{d} \approx 2000~u^{1/6}_{rad} \label{eqn:eqA3} 
\end{equation}

At a distance R$_{out}$ from an isotropic point source of total luminosity L,

\begin{equation}
u_{rad} = \frac{L}{4\pi R^{2}_{out}c} \label{eqn:eqA4}
\end{equation}

so

\begin{equation}
R_{out} \approx (\frac{2000}{T_{min}})^{3} (4\pi c)^{-1/2} L^{1/2} \label{eqn:eqA5}
\end{equation}

In astronomically convenient units,

\begin{equation}
(\frac{R_{out}}{pc}) \approx (\frac{L}{2.3 \times 10^{44} erg s^{-1}})^{1/2} (\frac{400 K}{T_{min}})^{3} \label{eqn:eqA6}
\end{equation}

or

\begin{equation}
log (\frac{R_{out}}{pc}) \approx 0.5 [log(L) - 44.36] -3 log(\frac{T_{min}}{400 K}) \label{eqn:eqA7}
\end{equation}

Inserting $L = 2.3 \times 10^{44} erg s^{-1}$ and $R_{out} = 1 pc \approx 3.09 \times 10^{18} cm$
into Equation \ref{eqn:eqA4} gives $u_{rad} \sim 10^{-4} erg cm^{-3}$ and $U \sim 10^{8}$.

\clearpage

\begin{deluxetable}{llllllllll}
\rotate
\tablecolumns{10} \tablewidth{0pc} 
\tablecaption{Source Information}
\tablehead {
  \colhead{Source name} & \colhead{R.A. (J2000)} &  \colhead{Dec. (J2000)} &  \colhead{$\delta$ R.A.} &  \colhead{$\delta$ Dec.} &  \colhead{from} &  \colhead{V$_{LSR}$} &  \colhead{$\delta$ V} &  \colhead{AGN} &  \colhead{Hubble} 
\\
  \colhead{}       & \colhead{(J2000)} &  \colhead{(J2000)} & \colhead{(mas)} & \colhead{(mas)} & \colhead{} & \colhead{(km s$^{-1}$)} & \colhead{(km s$^{-1}$)} & \colhead{Type} & \colhead{Type}
\\
  \colhead{(1)}       & \colhead{(2)} &  \colhead{(3)} & \colhead{(4)} & \colhead{(5)} & \colhead{(6)} & \colhead{(7)} & \colhead{(8)} & \colhead{(9)} & \colhead{(10)}  
            }
\startdata		
  J0437+2456    	&  04:37:03.6840 	   	&  24:56:06.837      	&1.3  	&2  		&VLBI   		&4825  	&40 		& Sy 2  	& ---  \\
  				&  04:37:03.67			&  24:56:06.8		&500		&500		&NED		&		&		&		&	\\
  ESO 558$-$G009   	&  07:04:21.0113  		& -21:35:18.948  	&14   	&14	  	&VLA-B 		&7655  	&27  		& Sy 2  	& Sc  \\
  Mrk 1029      		&  02:17:03.566    		&  05:17:31.43     	&500   	&500   	&NED	  	&9067  	&32  		& Sy 2  	& ---  \\ 
  NGC 5495      		&  14:12:23.35    		& -27:06 29.20     	&300   	&300   	&VLA-BnA  	&6741  	& 9  		& Sy 2  	& SAB(r)c  \\
  NGC 1320     		&  03:24:48.70   		& -03:02:32.30   	&300   	&300   	&VLA 		&2649  	&16  		& Sy 2  	& Sa  \\
\enddata
\tablecomments {Positions used for data correlation except for J0437+2456. The position for J0437+2456 (first row) comes from 
the phase-referenced observation, which has an offset of ($191,36.7$)~mas (east,north) from the correlation position. Position for NGC 5495 comes from Kondratko et al. (2006).}
\label{table:positions}
\end{deluxetable}

\begin{deluxetable}{lllllccc}
\rotate
\tablecolumns{8} \tablewidth{0pc}  
\tablecaption{VLBI Observation Details}
\tablehead {
\colhead{Project Code} & \colhead{Date Observed} &  \colhead{Source Name} &  \colhead{Antennas used} &  \colhead{Beam Size} &  \colhead{Track Length} &  \colhead{Sensitivity} &  \colhead{Observing} 
\\
  \colhead{}       & \colhead{} &  \colhead{} & \colhead{} & \colhead{(mas$\times$mas, deg)} & \colhead{(hour)} & \colhead{} & \colhead{Strategy} 
\\
  \colhead{(1)}       & \colhead{(2)} &  \colhead{(3)} & \colhead{(4)} & \colhead{(5)} & \colhead{(6)} & \colhead{(7)} & \colhead{(8)}  
            }
\startdata 		
  BB278 K	 	&  2010 Nov 22 	&  J0437+2456   	&VLBA+GB+EB 	 	&0.90$\times$0.27, $-$17  		&12 		&1.10  		&Phase-ref. \\ 
  BB278 L	 	&  2010 Nov 28	 	&  J0437+2456   	&VLBA+GB+EB  		&0.69$\times$0.23, $-$13  		&12 		&0.62		&Self-cal.  \\ 
  BB321 X0	&  2014 Jan 03  	&  J0437+2456   	&VLBA+GB+VLA  		&1.28$\times$0.50, $-$14  		&6 		&0.70  		&Self-cal. \\ 
  BB321 X1	&  2014 Jan 04  	&  J0437+2456   	&VLBA+GB+VLA  		&0.98$\times$0.38, $-$16  		&6 		&0.60  		&Self-cal. \\ 
  BB321 X4	&  2014 Jan 24  	&  J0437+2456   	&VLBA+GB+EB+VLA  	&1.11$\times$0.32, $-$15 	 		&6 		&0.50  		&Self-cal. \\ 
  X0+X1+X4     & 				&  J0437+2456		&					&1.11$\times$0.37, $-$15  		& --- 		&0.34  		&Self-cal. \\  
  BB278 T	 	&  2011 Nov 01  	&  ESO 558$-$G009  	&VLBA+GB     			&1.30$\times$0.34, $-$10   	&8  		&1.65  		&Self-cal. \\ 
  BB278 X	 	&  2012 Jan 11   	&  ESO 558$-$G009  	&VLBA+GB     			&1.39$\times$0.46, $-$2   	&8  		&1.33  		&Self-cal. \\
  BB313AF	&  2012 Nov 18		&  ESO 558$-$G009     &VLBA+GB			&1.30$\times$0.41, $-$9 			&8 		& ---  		&Phase-ref. \\
  T+X		&				&  ESO 558$-$G009	&					&1.30$\times$0.39, $-$6  			& --- 		&1.10  		&Self-cal. \\  
  BB313 C	 	&  2012 Feb 26  	&  NGC 5495     	&VLBA+GB     			&1.49$\times$0.38, $-$11  		&6  		&2.10  		&Self-cal. \\
  BB313 AA 	&  2012 Oct 25   	&  MRK 1029    	&VLBA+GB  			&1.14$\times$0.49, $-$3  			&10 		&1.40 		 &Self-cal. \\   
  BB313 AB 	&  2012 Nov 07   	&  NGC 1320    	&VLBA+GB  			&1.20$\times$0.39, $-$8	  		&9 		&1.46  		&Self-cal. \\   
  
\enddata
\tablecomments {Sensitivity listed in Column 7 is in the unit of mJy Beam$^{-1}$ Chan$^{-1}$. BB278K did not include BR, and BB278X did not include MK. The track length includes two geodetic blocks of 1.5 hours in total. Because of the low declination, the track lengths for ESO 558$-$G009 and NGC 5495 are just 8 and 6 hours, respectively. The channel size used in the final VLBI maps are 100 kHz (J0437+2456), 125 kHz (ESO558$-$G009, NGC 5495, MRK 1029) and 62.5 kHz (NGC 1320).}
\label{table:observation}
\end{deluxetable}

\begin{deluxetable}{ccccccc}
\tablecolumns{7} \tablewidth{0pc}  
\tablecaption{Sample data for J0437+2456}
\tablehead {
\colhead{Velocity} & \colhead{R.A.} &  \colhead{$\delta$RA} &  \colhead{DEC} &  \colhead{$\delta$DEC} &  \colhead{Surface Brightness} &  \colhead{Noise}
\\
  \colhead{km~s$^{-1}$}       & \colhead{mas} &  \colhead{mas} & \colhead{mas} & \colhead{mas} & \colhead{mJy Beam$^{-1}$ Chan$^{-1}$} & \colhead{mJy Beam$^{-1}$ Chan$^{-1}$}
\\
  \colhead{(1)}       & \colhead{(2)} &  \colhead{(3)} & \colhead{(4)} & \colhead{(5)} & \colhead{(6)} & \colhead{(7)} 
            }
\startdata
4872.99 &    0.0537 &   0.01535 &    0.0527 &   0.03773 &     4.49 &   0.34 \\
4871.59 &    0.0213 &   0.01034 &    0.0337 &   0.02288 &     6.94 &   0.34 \\
4870.20 &    0.0150 &   0.00539 &    0.0470 &   0.01243 &    13.07 &   0.34 \\
4868.81 &    0.0091 &   0.00515 &    0.0623 &   0.01202 &    13.54 &   0.34 \\
4867.41 &    0.0365 &   0.00564 &    0.0165 &   0.01324 &    11.91 &   0.34 \\
4866.02 &    0.0280 &   0.00505 &    0.0347 &   0.01262 &    13.57 &   0.34 \\
4864.63 &    0.0253 &   0.00421 &    0.0329 &   0.01015 &    16.75 &   0.34 \\
4863.23 &    0.0228 &   0.00387 &    0.0327 &   0.00923 &    20.12 &   0.34 \\
4861.84 &    0.0177 &   0.00382 &    0.0225 &   0.00914 &    20.14 &   0.34 \\
4860.45 &    0.0220 &   0.00412 &    0.0253 &   0.00985 &    18.65 &   0.34 \\
4859.05 &    0.0275 &   0.00446 &    0.0184 &   0.00996 &    17.73 &   0.34 \\
4857.66 &    0.0150 &   0.00386 &    0.0297 &   0.00851 &    18.99 &   0.34 \\
4856.27 &    0.0091 &   0.00413 &    0.0339 &   0.00906 &    19.97 &   0.34 \\
4854.87 &    0.0199 &   0.00338 &    0.0195 &   0.00748 &    22.49 &   0.34 \\
4853.48 &    0.0130 &   0.00384 &    0.0404 &   0.00862 &    19.54 &   0.34 \\
4852.09 &    0.0116 &   0.00452 &    0.0218 &   0.01061 &    15.62 &   0.34 \\
4850.69 &    0.0236 &   0.00478 &    0.0038 &   0.01122 &    14.72 &   0.34 \\
4849.30 &    0.0106 &   0.00605 &    0.0403 &   0.01357 &    12.00 &   0.34 \\
4847.91 &    0.0204 &   0.00678 &    0.0068 &   0.01570 &    10.59 &   0.34 \\
4846.51 &    0.0061 &   0.00595 &    0.0085 &   0.01396 &    10.90 &   0.34 \\
4845.12 &    0.0210 &   0.00528 &   -0.0191 &   0.01257 &    12.77 &   0.34 \\
4843.73 &   -0.0053 &   0.00978 &    0.0138 &   0.02126 &     7.51 &   0.34 \\
4842.33 &   -0.0282 &   0.01365 &    0.0864 &   0.02858 &     5.33 &   0.34 \\
4840.94 &   -0.0075 &   0.01451 &    0.0192 &   0.03682 &     4.55 &   0.34 \\

\enddata
\tablecomments {
Sample data for J0437+2456. (This table will be available in its entirety in a machine-readable form in the online
journal. A portion is shown here for guidance regarding its form and content.)
}
\label{table:rawdata}
\end{deluxetable}

\begin{deluxetable}{cccccccc}
\rotate
\tablecolumns{8} \tablewidth{0pc}  
\tablecaption{The SMBH Masses and Basic Properties of the Maser Galaxies}
\tablehead {
\colhead{Name} & \colhead{Dist.} &  \colhead{SMBH mass} &  \colhead{Fitted Recession Velocity} &  \colhead{R$_{in}$} &  \colhead{R$_{out}$} &  \colhead{P.A.} &  \colhead{Incl.}  
\\
  \colhead{}       & \colhead{(Mpc)} &  \colhead{(M$_{\odot}$)} & \colhead{(km~s$^{-1}$)} & \colhead{(pc)} & \colhead{(pc)} & \colhead{($^{\circ}$)} & \colhead{($^{\circ}$)}  
  \\
  \colhead{(1)}       & \colhead{(2)} &  \colhead{(3)} & \colhead{(4)} & \colhead{(5)} & \colhead{(6)} & \colhead{(7)} & \colhead{(8)}
            }
\startdata		
  J0437+2456    	&65.3 $\pm$ 3.6    	&2.9 $\pm$ 0.3 $\times~10^{6}$       	&4807.6 $\pm$ 10.5		&0.04$\pm$ 0.03 &0.13 $\pm$ 0.03  	&20 $\pm$ 2  	 &81 $\pm$ 1\\
  ESO 558$-$G009   	&107.6 $\pm$ 5.9 	&1.7 $\pm$ 0.1 $\times~10^{7}$	&7599.0 $\pm$ 14.0		&0.20$\pm$ 0.05 &0.47 $\pm$0.06  	&256 $\pm$ 2    &98 $\pm$ 1      \\
  NGC 5495      		&95.7   $\pm$ 5.3	&1.1 $\pm$ 0.2 $\times~10^{7}$	&6839.6 $\pm$ 67.0		&0.10 $\pm$ 0.05 &0.30 $\pm$ 0.05  &176 $\pm$ 5   &95 $\pm$ 1     \\
  Mrk 1029*      		&120.8 $\pm$ 6.6   	&1.9 $\pm$ 0.5 $\times~10^{6}$  	&9129.9 $\pm$ 26.0		&0.23 $\pm$ 0.12 &0.45 $\pm$ 0.12  &218 $\pm$ 10  &79 $\pm$ 2    \\
  NGC 1320*    		&34.2   $\pm$ 1.9	&5.5 $\pm$ 2.5 $\times~10^{6}$ 	&2829.2 $\pm$ 11.4		&--- 			    &--- 		  		&90 $\pm$ 10  		&90*  \\ 
\enddata
\tablecomments {Col. (1): Galaxy name. Col. (2): Distance, taken from NED using the Hubble flow with an $H_{0}$ of 73 km~s$^{-1}$ Mpc$^{-1}$. Col. (3): The fitted SMBH mass. Col. (4): Fitted recession velocity from VLBI. Col. (5-6): The inner and outer radius of the maser disk, estimated from the high-velocity maser features. Please refer to Sect. 4.2 for a more comprehensive analysis. Col. (7): Position angle of the disk. Col. (8): Inclination angle of the disk. We note that since the maser configuration in Mrk 1029 and NGC 1320 are not  typical for disk maser, their fitted SMBH mass, recession velocity, disk size and position angle listed here are just indicative under a thin disk model. And we cannot estimate the disk inclination angle for NGC 1320 due to the lack of systemic masers, so we fixed it to 90$^{\circ}$.}
\label{table:property}
\end{deluxetable}

\begin{deluxetable}{ccccccccc}
\rotate
\tablecolumns{9} \tablewidth{0pc}  
\tablecaption{The maser disk size, SMBH mass and luminosity from the ``clean" disk sample}
\tablehead {
\colhead{Name} & \colhead{SMBH mass} &  \colhead{Dist.} &  \colhead{R$_{in}$} &  \colhead{R$_{out}$} & \colhead{Log L$_{WISE^{*}}$} & \colhead{Log L$_{WISE 5.5"}$} & \colhead{Log L$_{[O III]}$} & \colhead{References}
\\
    \colhead{}       & \colhead{(10$^{7}M_{\odot}$)} &  \colhead{(Mpc)}  & \colhead{(pc)} & \colhead{(pc)}  & \colhead{(erg s$^{-1}$)} & \colhead{(erg s$^{-1}$)} & \colhead{(erg s$^{-1}$)} & \colhead{(for Col. 2, 8)} 
\\  
    \colhead{(1)}       &  \colhead{(2)}       &  \colhead{(3)}       &  \colhead{(4)}       &  \colhead{(5)}       &  \colhead{(6)}       &  \colhead{(7)}       &  \colhead{(8)} &  \colhead{(9)}       
                }
\startdata		
NGC 4258 		& 4.00 $\pm$ 0.09 	&7.60		& 0.11 $\pm$ 0.004			& 0.29 $\pm$ 0.01  			&40.87	&40.94	&38.77	& 3,  10 \\ 
NGC 4388		& 0.84 $\pm$ 0.09	&19.0 		& 0.24 $\pm$ 0.03			& *0.29$^{+0.27}_{-0.03}$		&42.00	&42.08	&40.40	& 1,  9 \\ 
NGC 2273		& 0.75 $\pm$ 0.05	&25.7		& 0.03 $\pm$ 0.01			& *0.084$^{+0.282}_{-0.013}$	&42.19	&42.26	&40.50	& 1,  9 \\ 
IC 2560			& 0.44 $\pm$ 0.09	&44.5		& 0.11 $\pm$ 0.02			& 0.47 $\pm$ 0.03			&42.44	&42.50	&41.00	& 8,  9 \\ 
UGC 3789		& 1.04 $\pm$ 0.06	&45.4		& 0.08 $\pm$ 0.02			& *0.30$^{+0.19}_{-0.03}$		&42.07	&42.16	&41.30	& 2,  9 \\ 
NGC 1194			& 6.5	   $\pm$ 0.4 	&53.2		& *0.54 $^{+0.03}_{-0.25}$	& *1.33$^{+0.97}_{-0.06}$		&42.61	&42.68	&39.90	& 1,  9 \\ 
NGC 3393		& 3.10 $\pm$ 0.37	&56.2		& 0.17 $\pm$ 0.02			& *0.60$^{+0.80}_{-0.04}$		&42.37	&42.44	&41.30	& 7,  9 \\ 
J0437+2456		& 0.29 $\pm$ 0.03	&65.3		& 0.06 $\pm$ 0.03			& 0.13 $\pm$ 0.03			&41.56	&41.65	&39.00	& 5,  11 \\ 
NGC 2960		& 1.16 $\pm$ 0.07	&72.2		& 0.13 $\pm$ 0.04			& *0.37$^{+0.71}_{-0.04}$		&42.29	&42.38	&40.50	& 1,  9 \\ 
NGC 5495		& 1.05 $\pm$ 0.20	&95.7		& 0.10 $\pm$ 0.05			& 0.30 $\pm$ 0.05			&42.23	&42.36	&39.80	& 5,  12 \\ 
NGC 6323		& 0.94 $\pm$ 0.04	&106.0		& 0.13 $\pm$ 0.05			& 0.30 $\pm$ 0.05			&42.11	&42.22	&40.40	& 1,  9 \\ 
ESO 558$-$G009	& 1.70 $\pm$ 0.14	&107.6		& 0.20 $\pm$ 0.05			& 0.47 $\pm$ 0.06			&42.38	&42.46	& ---		& 5,  --- \\ 
NGC 5765b		& 4.55 $\pm$ 0.31	&117.0		& 0.30 $\pm$ 0.06			& 1.15 $\pm$ 0.07			&43.16	&43.24	&41.10	& 4,  11 \\ 
NGC 6264		& 2.91 $\pm$ 0.11	&139.4		& 0.24 $\pm$ 0.07			& 0.80 $\pm$ 0.07			&42.23	&42.29	&41.30	& 1,  9 \\ 
UGC 6093		& 2.65 $\pm$ 0.23	&153.2		& 0.12 $\pm$ 0.07			& *0.24$^{+0.29}_{-0.08}$		&41.88	&42.01	&40.20	& 6,  11 \\ 
\enddata
\tablecomments {
Col. (1): Galaxy name. Col. (2): SMBH masses. We have added systematic errors to the uncertainties of NGC 3393, IC 2560 and UGC 6093. See the text for more details. Col. (3): Distances, taken from NED using the Hubble flow, except for NGC 4258 and NGC 4388. The distance for NGC 4258 is adopted from Humphreys et al. 2013. We adopted the distance for NGC 4388 from Kuo et al. (2011). Col. (4): Maser disk inner radius. Col. (5): Maser disk outer radius. The maser disk size comes from Table~\ref{table:radiidata}. Radius starts with an asterisk indicates the radius estimated from VLBI are not consistent with results from single-dish spectrum, within 3$\sigma$, so we use the VLBI radius as the final value, with 1$\sigma$ uncertainty covering the difference between VLBI and single-dish measured results. Col. (6): 12 $\mu$m luminosity measured from the central pixel on the WISE Band 3 image. Col. (7): 12 $\mu$m luminosity measured from 5.5" aperture photometry reported from WISE online catalog. Col. (8): Optical O[III] luminosity from literature. References. (1) Kuo et al. 2011; (2) Reid et al. 2013; (3) Humphreys et al. 2013; (4) Gao et al. 2016; (5) this work; (6) Zhao et al. in prep; (7) Kondratko et al. 2008; (8) Wagner et al. in prep; (9) Greene et al. 2010; (10) Ho et al. 1997; (11) SDSS online catalog; (12) Greene J. E. priv. comm.
}
\label{table:diskdata}
\end{deluxetable}

\begin{deluxetable}{ccccccccc}
\rotate
\tablecolumns{9} \tablewidth{0pc}  
\tablecaption{The maser disk size measured with different methods from the ``clean" disk sample}
\tablehead {
\colhead{Name}  &  \colhead{V$_{sys}$} &  \colhead{R$_{in}^{V-sys}$} &  \colhead{R$_{in}^{V-hi}$} &  \colhead{R$_{in}^{S}$} &  \colhead{R$_{out}^{V-sys}$} &  \colhead{R$_{out}^{V-hi}$} &  \colhead{R$_{out}^{S}$} & \colhead{References}
\\
    \colhead{}     &  \colhead{(km s$^{-1}$)} &  \colhead{(pc)} & \colhead{(pc)} & \colhead{(pc)}  & \colhead{(pc)} & \colhead{(pc)}  & \colhead{(pc)} & \colhead{} 
\\  
    \colhead{(1)}       &  \colhead{(2)}       &  \colhead{(3)}       &  \colhead{(4)}       &  \colhead{(5)}  &  \colhead{(6)}  &  \colhead{(7)} &  \colhead{(8)}  &  \colhead{(9)}             
            }
\startdata		
NGC 4258 		& 474	& 0.14 $\pm$ 0.002	& 0.11 $\pm$ 0.004		& 0.12 $\pm$ 0.003	& 0.15 $\pm$ 0.002	& 0.29 $\pm$ 0.01		& 0.31 $\pm$ 0.01	& 3 \\ 
NGC 4388		& 2527	& ---				& 0.24 $\pm$ 0.03		& 0.21 $\pm$ 0.02	& ---				& 0.29 $\pm$ 0.03		& 0.56 $\pm$ 0.06	& 1 \\   
NGC 2273		& 1832	& ---				& 0.03 $\pm$ 0.01		& 0.03 $\pm$ 0.002	& ---				& 0.08 $\pm$ 0.01		& 0.37 $\pm$ 0.04	& 1 \\    
IC 2560			& 2906	& 0.08 $\pm$ 0.006	& 0.11 $\pm$ 0.02		& 0.09 $\pm$ 0.01	& 0.087$\pm$ 0.01	& 0.47 $\pm$ 0.03		& 0.54 $\pm$ 0.08 	& 8 \\    
UGC 3789		& 3262	& 0.08 $\pm$ 0.002	& 0.08 $\pm$ 0.02		& 0.06 $\pm$ 0.004	& 0.22 $\pm$ 0.01 	& 0.30 $\pm$ 0.03		& 0.49 $\pm$ 0.06	& 2 \\   
NGC 1194			& 4051	& 0.44 $\pm$ 0.01	& 0.54 $\pm$ 0.03 		& 0.29 $\pm$ 0.02	& 1.38 $\pm$ 0.04	& 1.33 $\pm$ 0.06		& 2.30 $\pm$ 0.02	& 1 \\    
NGC 3393		& 3750	& 0.16 $\pm$ 0.01	& 0.36 $\pm$ 0.02		& 0.30 $\pm$ 0.02	& 0.176$\pm$0.01	& 0.64 $\pm$ 0.04		& 1.40 $\pm$ 0.11	& 7 \\   
J0437+2456		& 4807	& 0.07 $\pm$0.003	& 0.04 $\pm$ 0.03 		&0.06 $\pm$ 0.01	& 0.126$\pm$0.01	& 0.13 $\pm$ 0.03		& 0.15 $\pm$ 0.02	& 5 \\    
NGC 2960		& 4945	& ---				& 0.13 $\pm$ 0.04		& 0.11 $\pm$ 0.01	& ---				& 0.37 $\pm$ 0.04		& 1.08 $\pm$ 0.16	& 1 \\    
NGC 5495		& 6839	& ---				& 0.10 $\pm$ 0.05		& 0.08 $\pm$0.02	& ---				& 0.30 $\pm$ 0.05		& 0.48 $\pm$ 0.23	& 5 \\ 
NGC 6323		& 7848	& 0.16 $\pm$ 0.003	& 0.13 $\pm$ 0.05		& 0.12 $\pm$ 0.01	& 0.40 $\pm$ 0.01	& 0.30 $\pm$ 0.05		& 0.35 $\pm$ 0.03	& 1 \\ 
ESO 558$-$G009	& 7597	& 0.20 $\pm$0.01	& 0.20 $\pm$ 0.05		& 0.21 $\pm$ 0.02	& 0.341$\pm$0.01	& 0.47 $\pm$ 0.06		& 0.49 $\pm$ 0.05	& 5 \\ 
NGC 5765b		& 8335	& 0.39 $\pm$0.01	& 0.30 $\pm$ 0.06		& 0.29 $\pm$ 0.02	& 0.88 $\pm$ 0.03	& 1.15 $\pm$ 0.07 		& 1.29 $\pm$ 0.09	& 4 \\ 
NGC 6264		& 10213	& 0.18 $\pm$0.003	& 0.24 $\pm$ 0.07		& 0.20 $\pm$ 0.01	& 0.425$\pm$0.01	& 0.80 $\pm$ 0.07		& 0.99 $\pm$ 0.09	& 1 \\ 
UGC 6093		& 10828	& ---				& 0.12 $\pm$ 0.07		& 0.11 $\pm$ 0.01	& ---				& 0.24 $\pm$ 0.08		& 0.53 $\pm$ 0.04	& 6 \\ 
\enddata
\tablecomments {
Estimation of the maser disk radii from both single dish spectrum and VLBI map. When possible, we also estimate the range of radii that systemic masers span by either using the acceleration data or slope on the position-velocity diagram. Refer to Section 4.2 for details. Col. (1): galaxy name. Col. (2): recession velocity derived from VLBI. Col. (3): the inner radius calculated from systemic masers from VLBI. Col. (4): the inner radius calculated from high-velocity masers in VLBI. Col. (5): the inner radius calculated from single-dish spectrum. Col. (6): the outer radius calculated from systemic masers from VLBI. Col. (7): the outer radius calculated from high-velocity masers in VLBI. Col. (8): the outer radius calculated from single-dish spectrum.  Col. (9) Reference for the VLBI results. References. (1) Kuo et al. 2011; (2) Reid et al. 2013; (3) Humphreys et al. 2013; (4) Gao et al. 2016; (5) this work; (6) Zhao et al. in prep; (7) Kondratko et al. 2008; (8) Wagner et al. in prep; 
}
\label{table:radiidata}
\end{deluxetable}


\clearpage
\begin{thebibliography}{99}
\bibitem[Braatz et al. (2010)]{Braatz:10}
    Braatz, J. A., Reid, M. J., Humphreys, E. M. L., et al. 2010, \apj, 718, 657
\bibitem[Braatz et al. (in prep)]{Braatz:15}
    Braatz, J. A. et al. in preparation
\bibitem[Collison & Watson (1995)]{Collison:95}
    Collison, A. J. \& Watson, W. D., 1995, \apj, 452, L103
\bibitem[de Vaucouleurs et al. (1991)]{de Vaucouleurs:91} 
de Vaucouleurs, G., de Vaucouleurs, A., Corwin J. R., et al. 1991, RC 3.9
\bibitem[Deller (2007)]{Deller:07}
    Deller, A. T., Tingay, S. J., Bailes, M. \& West, C., 2007, \pasp, 119, 318
\bibitem[Draine (2011)]{Draine:11}
    Draine, B. T., 2011, $Physics of the Interstellar and Intergalactic Medium$, 
    Princeton University Press. ISBN: 978-0-691-12214-4 
\bibitem[Draine \& Li (2007)]{Draine:07}
    Draine, B. T. \& Li, A., 2007, \apj, 657, 810 
\bibitem[Fu et al. (2010)]{Fu:10}
    Fu, H., Yan, L., Scoville, N. Z., et al. 2010, \apj, 722, 653 
\bibitem[Gammie et al. (1999)]{Gammie:99}
    Gammie, C. F., Narayan, R. \& Blandford, R., 1999, \apj, 516, 177   
\bibitem[Gao et al. (2016)]{Gao:16}
    Gao, F., Braatz, J. A., Reid, M. J., et al. 2016, \apj, 817, 128
\bibitem[Greene et al. (2010)]{Greene:10}
    Greene, J. E., Peng, C. Y., Kim, M., et al. 2010, \apj, 721, 26
\bibitem[Greene et al. (2016)]{Greene:16}
    Greene, J. E., Seth, A., Kim, M., et al. 2016, \apj, accepted
\bibitem[Greenhill et al. (1995)]{Greenhill:95}
    Greenhill, L. J., Jiang, D. R., Moran, J. M., et al. 1995, \apj, 440, 619    
\bibitem[Greenhill \& Gwinn (1997)]{Greenhill:97}
    Greenhill, L. J. \& Gwinn, C. R. 1997, \apss, 248, 261
\bibitem[Greenhill et al. (2003)]{Greenhill:03}
    Greenhill, L. J., Booth, R. S., Ellingsen, S. P., et al. 2003, \apj, 590, 162
\bibitem[Greenhill et al. (2009)]{Greenhill:09}
    Greenhill, L. J., Kondratko, P. T., Moran, J. M. \& Tilak, A. 2009, \apj, 707, 787
\bibitem[Heckman et al. (2004)]{Heckman:04}
    Heckman, T. M., Kauffmann, G., Brinchmann, J., et al. 2004, \apj, 613, 109
\bibitem[Herrnstein et al. (2005)]{Herrnstein:05}
    Herrnstein, J. R., Moran, J. M., Greenhill, L. J. \& Trotter, A. S. 2005, \apj, 629, 719
\bibitem[Ho et al. (1997)]{Ho:97}
    Ho, L. C., Filippenk, A. V. \& Sargent, W. L. W. 1997, \apjs, 112, 315
\bibitem[Huchra et al. (1999)]{Huchra:99}
    Huchra, J. P., Vogeley, M. S., \& Geller, M. J., 1999, \apjs, 121, 287    
\bibitem[Humphreys et al. (2013)]{Humphreys:13}
    Humphreys, E. M. L., Reid, M. J., Moran, J. M., Greenhill, L. J. \& Argon, A. L. 2013, \apj, 775,13
\bibitem[Kondratko et al. (2006)]{Kondratko:06}
   Kondratko, P. T., Greenhill, L. J., Moran, J. M. et al. 2006, \apj, 638, 100
\bibitem[Kondratko et al. (2008)]{Kondratko:08}
   Kondratko, P. T., Greenhill, L. J., Moran, J. M. et al. 2008, \apj, 678, 87
\bibitem[Kormendy \& Ho (2013)]{Kormendy:13}
   Kormendy, J. \& Ho, L. C., 2013, ARA\&A, 51,511
\bibitem[Kuo et al. (2011)]{Kuo:11}
    Kuo, C. Y., Braatz, J. A., Condon, J. J. et al. 2011, \apj, 727, 20
 \bibitem[Liu et al. (2009)]{Liu:09}
    Liu, X., Zakamska, N. L., Greene, J. E., et al. 2009, \apj, 702, 1098
 \bibitem[Masini et al. (2016)]{Masini:16}
    Masini, A., Comastri, A., Balokovic, M., et al. 2016, A\&A, arXiv160203185M
 \bibitem[Masters et al. (2006)]{Masters:06}
    Masters, K. L., Springob, C. M., Haynes, M. P. \& Giovanelli, R., 2006, \apj, 653, 861     
 \bibitem[Mateos et al. (2012)]{Mateos:12}
    Mateos, S., Alonso-Herrero, A., Carrera, F. J., et al. 2012, \mnras, 426, 3271    
\bibitem[Mioduszewski et al. (2000)]{Mioduszewski:00}
    Mioduszewski, A. \& Kogan, L. 2000, AIPS Memo 110
\bibitem[Miyoshi et al. (1995)]{Miyoshi:95}
    Miyoshi, M., Moran, J. M., Herrnstein, J. R., et al. 1995, \nat, 373,127     
\bibitem[Neufeld et al. (1994)]{Neufeld:94}
    Neufeld, D. A., Maloney, P. R. \& Conger, S. 1994, \apj, 436, L127   
\bibitem[Neufeld & Maloney (1995)]{Neufeld:95}
    Neufeld, D. A. \& Maloney, P, R., 1995, \apj, 447, L17 
\bibitem[Pesce et al. (2015)]{Pesce:15}
    Pesce, D. W., Braatz, J. A., Condon, J. J. et al. 2015, \apj, 810, 65
\bibitem[Planck (2015)]{Planck:15}
    Planck Collaboration, Ade, P. A. R., Aghanim, N., et al. 2015, arXiv:1502.01589
\bibitem[Reid et al. (2009)]{Reid:09}
    Reid, M. J., Braatz, J. A., Condon, J. J. et al. 2009, \apj, 695, 287
\bibitem[Reid et al. (2013)]{Reid:13}
    Reid, M. J., Braatz, J. A., Condon, J. J. et al. 2013, \apj, 767, 154
\bibitem[Riess et al. (2016)]{Riess:16}
    Riess, A. G., Macri, L. M., Hoffmann, S. L. et al. 2016, arXiv:1604.01424
\bibitem[Stern et al. (2012)]{Stern:12}
    Stern, D., Assef, R. J., Benford, D. J., et al. 2012, \apj, 753, 30 
\bibitem[Thompson et al. (2001)]{Thompson:01}
    Thompson, A. R., Moran, J. M., \& Swenson, G. W., Jr. 2001, Interferometry and Synthesis in Radio Astronomy (New York: Wiley) 
\bibitem[Bosch et al. (2016)]{Bosch:16}
  van den Bosch, R. C. E., Greene, J. E., Braatz, J. A., Constantin, A., \& Kuo, C. Y., 2016, \apj, 819, 11
\bibitem[Wagner et al. (in prep)]{Wagner:15}
  Wagner, J. et al. in preparation
\bibitem[Wardle \& Yusef-Zadeh (2012)]{Wardle:12}
    Wardle, M. \& Yusef-Zadeh, F. 2012, \apj, 750, L38	
\bibitem[Wright et al. (2010)]{Wright:10}
    Wright, E. L., Eisenhardt, P. R. M., Mainzer, A. K., et al. 2010, \apj, 140, 1868
\bibitem[Zakamska et al. (2003)]{Zakamska:03}
    Zakamska, N. L., Strauss, M. A., Krolik, J. H., et al. 2003, \aj, 126, 2125
\bibitem[Zhao et al. (in prep)]{Zhao:15}
    Zhao, W. et al. in preparation
\bibitem[Zhang et al. (2006)]{Zhang:06}
    Zhang, J. S., Henkel, C., Kadler, M., et al. 2006, A\&A, 450, 933
\bibitem[Zhang et al. (2010)]{Zhang:10}
    Zhang, J. S., Henkel, C., Guo, Q., Wang, H. G., \& Fan, J. H., 2010, \apj, 708, 1528
\bibitem[Zhang et al. (2012)]{Zhang:12}
    Zhang, J. S., Henkel, C., Guo, Q., \& Wang, J., 2012, A\&A, 538, 152
\end{thebibliography}
\end{document}